\pdfoutput=1

\documentclass[11pt,a4paper]{article}
\usepackage{jheppub}
\usepackage{amssymb}
\usepackage{amsthm}
\usepackage{amstext}
\usepackage{amsmath}
\usepackage{amsfonts}
\usepackage{verbatim}
\usepackage{latexsym}
\usepackage{t1enc}
\usepackage{graphicx}
\usepackage[all]{xy}
\usepackage{makeidx}
\usepackage{slashed}
\usepackage{multicol}
\UseRawInputEncoding
 
\numberwithin{equation}{section}

\def\l{\langle}
\def\r{\rangle}
\def\be{\begin{equation}}
\def\ee{\end{equation}}
\def\ba{\begin{eqnarray}}
\def\ea{\end{eqnarray}}
\def\nl{\nonumber\\}
\def\Li{\mbox{Li}}
\def\Res{\mbox{Res}}

\title{Uniqueness of Two-Loop Master Contours}
\author[a]{Simon Caron-Huot,}
\author[a,b,c]{Kasper J. Larsen}
\affiliation[a]{School of Natural Sciences, Institute for Advanced Study,\\
Princeton, NJ 08540, USA}
\affiliation[b]{Department of Physics and Astronomy, Uppsala University,\\
SE-75108 Uppsala, Sweden}
\affiliation[c]{Institut de Physique Théorique, CEA-Saclay,\\
F-91191 Gif-sur-Yvette cedex, France} \emailAdd{schuot@ias.edu}
\emailAdd{kasper.larsen@cea.fr}
\abstract{
Generalized-unitarity calculations of two-loop amplitudes
are performed by expanding the amplitude in a basis of master
integrals and then determining the coefficients by taking a number
of generalized cuts. In this paper, we present a complete classification
of the solutions to the maximal cut of integrals with the double-box topology.
The ideas presented here are expected to be relevant for all two-loop topologies as well.
We find that these maximal-cut solutions are naturally associated with Riemann surfaces
whose topology is determined by the number of states at the vertices
of the double-box graph. In the case of four massless external momenta we find that,
once the geometry of these Riemann surfaces is properly understood,
there are uniquely defined master contours producing the coefficients
of the double-box integrals in the basis decomposition of the two-loop amplitude.
This is in perfect analogy with the situation in one-loop generalized unitarity.

In addition, we point out that the chiral integrals recently introduced
by Arkani-Hamed et al. can be used as master integrals for the double-box
contributions to the two-loop amplitudes in any gauge theory.
The infrared finiteness of these integrals allow for their
coefficients as well as their integrated expressions to
be evaluated in strictly four dimensions, providing significant technical simplification.
We evaluate these
integrals at four points and obtain remarkably compact results.}


\begin{document}

\begin{flushright}\begin{tabular}{r}
Saclay IPhT--T12/028
\\ UUITP-13/12
\end{tabular}\end{flushright}
\vspace{-14.1mm}
\maketitle

\pagebreak

\section{Introduction}

The study of scattering amplitude in gauge theories is a fascinating
subject, partly because of the close link it provides between
theory and experiment, and partly because it continues to unravel
interesting structures in quantum field theory.
Our understanding
of computing amplitudes has undergone a revolution over the past
decade and a half, owing in large part to the development of
on-shell recursion relations for tree-level amplitudes
\cite{Britto:2004ap,Britto:2005fq} and a purely on-shell formalism
for loop-level amplitudes, the modern unitarity method
\cite{Bern:1994zx,Bern:1994cg,Bern:1996je,Bern:1995db,Bern:1997sc,Bern:1996ja,
Britto:2004nc,Britto:2004nj,Bidder:2004tx,Bidder:2005ri,Bidder:2005in,Bern:2005hh,
BjerrumBohr:2007vu,Bern:2005cq,Britto:2005ha,Britto:2006sj,Mastrolia:2006ki,
Brandhuber:2005jw,Ossola:2006us,Bern:2007dw,Forde:2007mi,Badger:2008cm,
Anastasiou:2006jv,Anastasiou:2006gt,Giele:2008ve,Britto:2006fc,Britto:2007tt,
Britto:2008vq,Britto:2008sw,Berger:2009zb,Bern:2010qa,Kosower:2011ty,Larsen:2012sx}.
These powerful modern methods have to a large extent replaced the more
traditional Feynman diagrammatic approach for tree-level
and one-loop amplitudes. Thus, the current frontier is the
development of systematic approaches for computing two-loop
amplitudes.

Impressive calculations of two-loop amplitudes have been done
by means of Feynman diagrams, including all parton-level amplitudes
required for $e^+ e^-$ annihilation into three jets, to give just one example
\cite{GehrmannDeRidder:2007jk,Weinzierl:2008iv}
(these computations were subsequently used to extract $\alpha_s$ to $1\%$ accuracy from the
three-jet LEP data \cite{Dissertori:2009ik,Dissertori:2009qa}).
The focus of this paper is to explore a different approach, the unitarity method at two loops. This method has proven
very successful at one loop where it has rendered a number of
amplitude calculations possible, in particular of processes with many partons
in the final state. In this formalism, the one-loop amplitude
is written as a sum over a set of basis integrals, with coefficients that are rational
in external spinors,
\begin{equation}
{\rm Amplitude} = \sum_{j\in {\rm Basis}}
  {\rm coefficient}_j {\rm Integral}_j +
{\rm Rational} \: . \label{BasicEquation}
\end{equation}
The process-dependence thus resides in the integral coefficients which
are the object of calculation within the unitarity-based approach. The determination
of the coefficients is done
by applying to both sides of this basis decomposition
a number of cuts, defined in the basic variant of unitarity as computing the
branch cut discontinuities across the various kinematic channels. As a result,
the left hand side of eq.~(\ref{BasicEquation}) is, by the Cutkosky rules, turned into
a product of tree-level amplitudes, enabling the computation of one-loop
amplitudes from tree-level data.

One-loop unitarity also exists in a more recent version, called generalized
unitarity, in which the operation of taking cuts does not have
any known interpretation in terms of branch cut discontinuities. Rather,
generalized cuts are defined as a change of the integration range
away from the real slice $\mathbb{R}^D$ (where $D=4-2\epsilon$) into
a contour of real dimension 4, embedded in $\mathbb{C}^4$.
The resulting contour integrations compute residues that are unique to each of the
basis integrals in eq.~(\ref{BasicEquation}), enabling a direct extraction of their coefficients,
using only tree-level amplitudes as input data.

Unitarity has also been applied beyond one loop, taking as the
starting point a decomposition, similar to eq.~(\ref{BasicEquation}), of the desired amplitude into a
(typically overcomplete) basis and requiring agreement between
the two sides on all cuts to determine the integral coefficients
on the right hand side. Several impressive
calculations have been done in this way, primarily in $\mathcal{N}=4$ supersymmetric
Yang-Mills theory and in $\mathcal{N}=8$ supergravity. However,
calculations of this nature require crafty choices of bases, and it
is fair to say that no systematized use of generalized
unitarity exists beyond one loop.

In ref.~\cite{Kosower:2011ty}, the first steps were taken in
developing a fully systematic version of generalized unitarity
at two loops. In the approach followed there,
the two-loop amplitude is decomposed as a linear combination of basis
integrals, in similarity with eq.~(\ref{BasicEquation}). The integral
coefficients are determined by applying to both sides of this two-loop basis decomposition
so-called augmented heptacuts, defined as a change of the integration range
away from the real slice $\mathbb{R}^D \times \mathbb{R}^D$
into contours of real dimension 8, embedded in $\mathbb{C}^8$.
These contours are particular linear combinations of the tori encircling
the leading singularities of the integrand, and satisfying the consistency condition that
any function which integrates to zero on the real slice
$\mathbb{R}^D \times \mathbb{R}^D$ (where $D=4-2\epsilon$)
must also integrate to zero on the $\mathbb{C}^8$-embedded contour.
This constraint on the contour ensures that two Feynman integrals
which are equal, possibly through some non-trivial relations, will
also have identical maximal cuts.
As explained in ref.~\cite{Kosower:2011ty},
contours satisfying this consistency condition, or \emph{master contours}, are guaranteed to produce correct results
for scattering amplitudes in any gauge theory.
A closely related, but distinct, approach is that of refs.~\cite{Mastrolia:2011pr,Badger:2012dp},
in which the heptacut integrand is reconstructed by polynomial matching in similarity
with the OPP approach \cite{Ossola:2006us}.

A perplexing feature of the contours obtained in ref.~\cite{Kosower:2011ty} is that they were not
found to be unique,
in contradistinction with the situation found at one loop \cite{Britto:2004nc,Forde:2007mi}.
In this paper, we will extend the results of ref.~\cite{Kosower:2011ty} to an arbitrary number of external legs, and at the same time
show that the contours are actually unique, once proper identifications of the leading singularity cycles are taken into account.
In this way, the situation at two loops becomes entirely analogous to the situation at one loop.
\\
\\
This paper is organized as follows. In section~\ref{sec:max_cut_of_DB}, we introduce
notation and formulas used throughout the paper. In section~\ref{sec:solutions}, we
give our classification of the kinematical solutions to the
maximal cut constraints of the general double-box integral and
discuss the singularities of the Jacobian arising from linearizing
these constraints. In section~\ref{sec:twistors}, we explain how to rephrase the
problem of solving on-shell constraints as a geometric problem in momentum
twistor space. In section~\ref{sec:proliferation_of_contours}, we prove that the master contours
extracting the double-box coefficients of two-loop amplitudes are
uniquely defined, once the sharing of Jacobian poles between
kinematical solutions is properly taken into account. We then show
that the double-box-topology basis elements can be chosen to have
chiral numerator insertions. In section~\ref{sec:analytical_evaluation_of_chiral_DBs},
we give a detailed derivation of the analytic expressions of these
chiral double-box integrals. In section~\ref{sec:conclusions}, we provide our
conclusions and suggest directions for future investigation.

\section{Maximal cut of the general double box}\label{sec:max_cut_of_DB}

Unitarity-based computations of two-loop
amplitudes take as their starting point
the expansion of the amplitude into a basis of linearly independent
two-loop integrals,
\begin{equation}
A^{(2)} \hspace{1mm}=\hspace{1mm} \sum_{i} c_i(\epsilon) \hspace{0.4mm} \mbox{Int}_i\:. \label{eq:2-loop_basis_decomposition}
\end{equation}
The form of the right hand side is obtained by applying integral
reductions to the Feynman-diagrammatic expansion of the amplitude, and the
integrals $\mbox{Int}_i$ are referred to as \emph{master integrals} (the rational
contributions will disregarded in this paper).
The process dependence resides in the integral coefficients $c_i$,
and the goal of unitarity calculations is to determine these
coefficients as functions of the external momenta.

\begin{figure}[!h]
\begin{center}
\includegraphics[angle=0, width=0.45\textwidth]{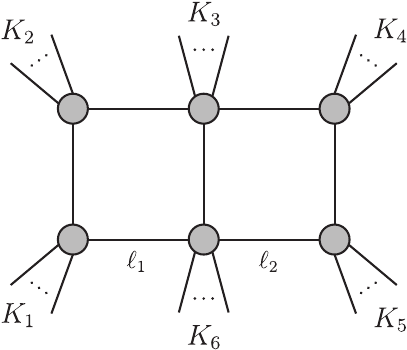}
\caption{The general double-box integral. The $\cdots$ dots at each vertex
represent the presence of an arbitrary number of massless legs.
Each of the vertices, shown as gray blobs, is given a label
$i=1,\ldots,6$ which equals the index of the associated external
momentum $K_i$.}\label{fig:general_double_box}
\end{center}
\end{figure}

In this paper, we will be concerned with the coefficients of the integrals in eq.~(\ref{eq:2-loop_basis_decomposition})
containing the maximal number of propagators.
These integrals turn out to have the double-box topology,\footnote{When the number of external
states exceeds four, the leading topology is that of a pentagon-box or a double-pentagon.
However, we expect the coefficients of such integrals to be simpler to extract due to the explicit octacuts they contain.}
illustrated in figure~\ref{fig:general_double_box}. They are defined by
\begin{eqnarray}
I_\mathrm{DB}[\Phi] \hspace{0.2mm}&\equiv&\hspace{0.2mm} \int \frac{d^D \ell_1}{(2 \pi)^D} \frac{d^D \ell_2}{(2 \pi)^D}
\left( \frac{\Phi (\ell_1, \ell_2)}{\ell_1^2 \hspace{0.6mm} (\ell_1-K_1)^2 \hspace{0.6mm} (\ell_1-K_1-K_2)^2 \hspace{0.6mm} (\ell_1+\ell_2+K_6)^2}
\right. \nonumber \\
&\phantom{=}& \hspace{40mm} \left. \times \frac{1}{\ell_2^2 \hspace{0.6mm} (\ell_2-K_5)^2 \hspace{0.6mm} (\ell_2-K_4-K_5)^2} \right)
\label{eq:def_double_box}
\end{eqnarray}
where $D=4-2\epsilon$. The function $\Phi (\ell_1, \ell_2)$ multiplied into
the integrand will be referred to as a \emph{numerator insertion}. The
special case of $\Phi = 1$ produces a so-called \emph{scalar} double-box integral.
The right hand side of eq.~(\ref{eq:2-loop_basis_decomposition})
will, depending on the number of external momenta in the process in question,
contain several integrals of the double-box topology
with various numerator insertions (for four external momenta, there are two master integrals;
the two used in ref.~\cite{Kosower:2011ty} involved the insertions $\Phi = 1$ and $\Phi = \ell_1 \cdot k_4$).

In order to determine the integral coefficients in eq.~(\ref{eq:2-loop_basis_decomposition}),
one applies to both sides a number of so-called generalized-unitarity cuts. These
can roughly speaking be understood as a replacement of the seven propagators
in eq.~(\ref{eq:def_double_box}) by seven $\delta$-functions whose arguments
are given as the corresponding inverse propagators. These $\delta$-functions
thereby solve what are called the on-shell constraints,
\begin{eqnarray}
                 \ell_1^2 &=& 0 \label{eq:on-shell_constraint_1}\\
         (\ell_1 - K_1)^2 &=& 0 \label{eq:on-shell_constraint_2}\\
   (\ell_1 - K_1 - K_2)^2 &=& 0 \label{eq:on-shell_constraint_3}\\
                 \ell_2^2 &=& 0 \label{eq:on-shell_constraint_4}\\
           (\ell_2-K_5)^2 &=& 0 \label{eq:on-shell_constraint_5}\\
   (\ell_2 - K_4 - K_5)^2 &=& 0 \label{eq:on-shell_constraint_6}\\
(\ell_1 + \ell_2 + K_6)^2 &=& 0 \label{eq:on-shell_constraint_7}\:,
\end{eqnarray}
whose solutions $\ell_1$, $\ell_2$ are generically complex.
The effect of applying such generalized cuts to eq.~(\ref{eq:2-loop_basis_decomposition})
is to turn the loop integrals into contour integrals in the complex
plane. Then (roughly speaking) by choosing the integration contours to encircle poles unique to each master
integral in this basis decomposition, one may extract their coefficients, thereby
determining the amplitude. We call such contours \emph{master contours}
and will discuss them in much greater detail in section~\ref{sec:proliferation_of_contours}.
\\
\\
To solve the on-shell constraints (\ref{eq:on-shell_constraint_1})-(\ref{eq:on-shell_constraint_7})
for an arbitary number of external momenta, it proves useful
to introduce, following refs.~\cite{Ossola:2006us,Forde:2007mi}, null
vectors $K_1^{\flat\mu}$ and $K_2^{\flat\mu}$ that lie within the
plane spanned by $K_1^\mu$ and $K_2^\mu$. Note that the vectors
$K_1^\mu$ and $K_2^\mu$ are not
necessarily assumed to be null, but the vectors
$K_1^{\flat \mu}$ and $K_2^{\flat \mu}$ are appropriate null
linear combinations. Similarly, we introduce vectors $K_4^{\flat}$
and $K_5^{\flat}$ in the plane spanned by $K_4$ and $K_5$.  Using
these vectors, a convenient parametrization of the loop momenta is
given by
\begin{eqnarray}
\ell_1^\mu \hspace{-1mm}&=&\hspace{-1mm} \alpha_1 K_1^{\flat \mu}
+ \alpha_2 K_2^{\flat \mu} + \alpha_3 \big\langle K_1^\flat |
\gamma^\mu | K_2^\flat \big] + \alpha_4 \big\langle K_2^\flat
 | \gamma^\mu | K_1^\flat \big] \label{eq:l1_parametrization} \\
\ell_2^\mu \hspace{-1mm}&=&\hspace{-1mm} \beta_1 K_4^{\flat \mu} +
\beta_2 K_5^{\flat \mu} + \beta_3 \big\langle K_4^\flat |
\gamma^\mu | K_5^\flat \big] + \beta_4 \big\langle K_5^\flat |
\gamma^\mu | K_4^\flat \big] \: . \label{eq:l2_parametrization}
\end{eqnarray}
Re-expressed in terms of the loop momentum parametrization (\ref{eq:l1_parametrization})-(\ref{eq:l2_parametrization}),
the on-shell constraints (\ref{eq:on-shell_constraint_1})-(\ref{eq:on-shell_constraint_6})
(corresponding to cutting the six outer propagators in figure~\ref{fig:general_double_box})
take the form
\begin{equation}
\begin{array}{lll}
\alpha_1 = \frac{\gamma_1(S_2 + \gamma_1)}{\gamma_1^2 - S_1 S_2}
\: , & \hspace{3mm} \alpha_2 = \frac{S_1 S_2 (S_1 +
\gamma_1)}{\gamma_1(S_1 S_2 - \gamma_1^2)} \: , & \hspace{3mm}
\alpha_3 \alpha_4 = -\frac{S_1 S_2 (S_1 + \gamma_1)(S_2 +
\gamma_1)}{4(\gamma_1^2 - S_1 S_2)^2} \\[2mm]
\beta_1 = \frac{S_4 S_5 (S_5 + \gamma_2)}{\gamma_2(S_4 S_5 -
\gamma_2^2)} \: , & \hspace{3mm} \beta_2 = \frac{\gamma_2 (S_4 +
\gamma_2)}{\gamma_2^2 - S_4 S_5} \: , & \hspace{3mm} \beta_3
\beta_4 = -\frac{S_4 S_5 (S_4 + \gamma_2)(S_5 +
\gamma_2)}{4(\gamma_2^2 - S_4 S_5)^2}
\end{array}\label{eq:onshell-values_alpha_beta}
\end{equation}
where $S_i\equiv K_i^2$, $\gamma_1 = K_1 \cdot K_2 \pm \sqrt{(K_1 \cdot K_2)^2
- K_1^2 K_2^2}$ and $\gamma_2$ is defined in analogy with $\gamma_1$.
We refer to refs.~\cite{Ossola:2006us,Forde:2007mi} for more details.
We see that in this parametrization, the variables
$\alpha_1,\alpha_2,\beta_1,\beta_2$ are directly fixed, while the
remaining variables obey simple constraints of the form
$\alpha_3\alpha_4=\textrm{constant}$. On the solution of the above
equations, cutting the central propagator gives rise to the
equation
\begin{eqnarray}
\hspace{-0.8cm}0=\frac12 (\ell_1 + \ell_2+K_6)^2 &=& \Big( \alpha_1
K_1^{\flat \mu} + \alpha_2 K_2^{\flat \mu} + \alpha_3 \big\langle
K_1^\flat | \gamma^\mu | K_2^\flat \big] + \alpha_4
\big\langle K_2^\flat | \gamma^\mu | K_1^\flat \big] + K_6^\mu \Big) \nonumber \\
&\phantom{=}& \hspace{-22mm} \times \hspace{0.5mm} \Big(\beta_1
K_{4\mu}^\flat + \beta_2 K_{5\mu}^\flat + \beta_3 \big\langle
K_4^\flat | \gamma_\mu | K_5^\flat \big] + \beta_4 \big\langle
K_5^\flat | \gamma_\mu | K_4^\flat \big] + K_{6\mu} \Big) -
\frac12K_6^2 \: .  \label{qsquared}
\end{eqnarray}
Cutting the seven propagators visible in the double-box graph in figure~\ref{fig:general_double_box}
will only fix seven out of the eight integration variables in the two-loop integration. Nevertheless,
as was explained in ref. \cite{Buchbinder:2005wp}, due to the Jacobian
factors arising from solving the $\delta$-functions, the measure
for the remaining variable can develop poles at specific
locations. The last integration can then be performed on a small circle
enclosing such poles, effectively pulling out an eighth
propagator. Such poles are referred to as the \emph{leading singularities} of
the integrand.
\\
\\
For future reference, let us provide a few details of the computation
of the maximal cut of the double-box integral in eq.~(\ref{eq:def_double_box}).
Replacing all propagators in eq.~(\ref{eq:def_double_box}) by $\delta$-functions
and integrating out the six that correspond to
the outer propagators in figure~\ref{fig:general_double_box}, the
heptacut measure on a given kinematical solution is
\begin{equation}
J = C \oint \frac{d\alpha_3 d\beta_3}{\alpha_3\beta_3}
\hspace{0.5mm} \hspace{0.5mm} \delta \big( (\ell_1 +
\ell_2+K_6)^2\big) \quad \mbox{where} \quad C\equiv\frac{\gamma_1
\gamma_2}{(\gamma_1^2-S_1S_2)(\gamma_2^2-S_4S_5)}
 \label{hexacut} \: .
\end{equation}
We note that the variables $\alpha_3$ or $\beta_3$ are not always good integration variables: on certain solutions to
eq.~(\ref{eq:onshell-values_alpha_beta}) they may happen to be constant.
In such cases, they should be traded for $\alpha_4$ or $\beta_4$ through $\frac{d\alpha_3}{\alpha_3}\to -\frac{d\alpha_4}{\alpha_4}$, and/or
$\frac{d\beta_3}{\beta_3} \to -\frac{d\beta_4}{\beta_4}$. Notice the relative signs, which are essential to ensure the global consistency of the residues.
These arise from the fact that, in solving for the $\delta$-functions, one should use determinants, not absolute values of determinants
(see, for instance, the discussion in ref. \cite{ArkaniHamed:2009dn}).

Integrating out the remaining $\delta$-function in eq.~(\ref{hexacut}), for each
kinematical solution, produces the corresponding maximal cut
\begin{equation}
J \hspace{1mm}=\hspace{1mm} C
\oint_{\Gamma} \frac{dz}{z} \Big( B_0(z)^2 - 4B_1 (z)
B_{-1}(z) \Big)^{-1/2} \label{eq:heptacut_Jacobian_S_2_and_S_6}
\end{equation}
where $z \equiv \alpha_3$ and
\begin{eqnarray}
B_1 &=& \big\langle K_4^\flat | \gamma_\mu | K_5^\flat \big] \Big(
\alpha_1 K_1^{\flat \mu} + \alpha_2 K_2^{\flat \mu} + z
\big\langle K_1^\flat | \gamma^\mu | K_2^\flat \big] + \alpha_4
(z) \big\langle K_2^\flat |
\gamma^\mu | K_1^\flat\big] + K_6^\mu \Big) \phantom{aaaaa} \label{eq:coeff_B_1} \\
B_0 &=& \Big( \beta_1 K_{4\mu}^\flat
+ \beta_2 K_{5\mu}^\flat + K_{6\mu} \Big) \nonumber \\
&\phantom{=}& \hspace{0.4cm} \times \hspace{0.5mm} \Big( \alpha_1
K_1^{\flat \mu} + \alpha_2 K_2^{\flat \mu} + z \big\langle
K_1^\flat | \gamma^\mu | K_2^\flat\big] + \alpha_4 (z) \big\langle
K_2^\flat | \gamma^\mu | K_1^\flat\big] + K_6^\mu \Big)
- \textstyle{\frac{1}{2}} K_6^2 \label{eq:coeff_B_0} \\
B_{-1} &=& -\frac{S_4 S_5(S_4 + \gamma_2)(S_5 + \gamma_2)
\big\langle K_5^\flat |\gamma_\mu |
K_4^\flat \big]}{4(\gamma_2^2 - S_4 S_5)^2} \nonumber \\
&\phantom{=}& \hspace{0.9cm} \times \hspace{0.5mm} \Big( \alpha_1
K_1^{\flat \mu} + \alpha_2 K_2^{\flat \mu} + z \big\langle
K_1^\flat | \gamma^\mu | K_2^\flat \big] + \alpha_4 (z)
\big\langle K_2^\flat | \gamma^\mu | K_1^\flat \big] + K_6^\mu
\Big) \label{eq:coeff_B_-1} \: .
\end{eqnarray}
Here $\alpha_1, \alpha_2, \alpha_4(z), \beta_1, \beta_2$ are given
by eq. (\ref{eq:onshell-values_alpha_beta}), whereas $z$ is
unconstrained, reflecting the degree of freedom left over after
imposing the seven cut constraints.  Similar formulas arise when
solving instead for $z=\alpha_4,\beta_3$ or $\beta_4$.

Despite appearances, we will see that in all cases with less than 10 massless particles,
the argument of the square root in eq.~(\ref{eq:heptacut_Jacobian_S_2_and_S_6}) is in fact a perfect square.
\\
\\
Our goal in the next section will be to determine when and where does the integrand of
eq.~(\ref{eq:heptacut_Jacobian_S_2_and_S_6}), referred to throughout
this paper as the (heptacut) Jacobian, give rise to poles.
As we will find, these poles are naturally associated with three-point vertices in the double-box graph
illustrated in figure~\ref{fig:general_double_box}, and their locations can be understood in a simple way.

\section{Kinematical solutions and Jacobian poles}\label{sec:solutions}

In this section we consider the classes of solutions to the joint
heptacut constraints (\ref{eq:on-shell_constraint_1})-(\ref{eq:on-shell_constraint_7}).
As the loop momenta have a total of eight
degrees of freedom $(\alpha_1, \ldots, \alpha_4, \beta_1, \ldots,
\beta_4)$, the result of imposing seven on-shell constraints will
be to fix all but one of these parameters. The various
choices of freezing the loop parameters to particular values that
solve these constraints span a number of distinct kinematical
solutions, whose unconstrained variable $z\in \mathbb{C}$
parametrizes a Riemann surface (for example, a Riemann sphere). As
we shall see, the Riemann surfaces associated with the kinematical
solutions are not disjoint, but rather they have pointwise
intersections located at the poles of the Jacobian discussed in
the previous section.

The number of kinematical solutions to the heptacut constraints is
determined by the distribution of external momenta at the vertices
of the double-box graph, and an important role in the classification is
played by the vertices that join three massless lines. In order to
state the classification, we introduce some notation which will be
used throughout the paper,
\begin{equation}
\begin{array}{llll}
N_i &\equiv& \mbox{\# of external legs at vertex $i$} &
\hspace{6mm} \mbox{for} \hspace{4mm} i=1,\ldots,6
\\[1.5mm]
n_i &\equiv& \mbox{total \# of legs at vertex $i$} & \hspace{6mm}
\mbox{for} \hspace{4mm} i=1,\ldots,6 \\[1.5mm]
\mu_j &\equiv& \left\{ \begin{array}{l} (1 - \delta_{n_1
\hspace{-0.2mm},\hspace{0.3mm} 3})(1 -
\delta_{n_2 \hspace{-0.2mm}, \hspace{0.3mm} 3}) \\
(1 - \delta_{n_3 \hspace{-0.2mm},\hspace{0.3mm} 3})(1 -
\delta_{n_6 \hspace{-0.2mm}, \hspace{0.3mm} 3}) \\
(1 - \delta_{n_4 \hspace{-0.2mm},\hspace{0.3mm} 3})(1 -
\delta_{n_5 \hspace{-0.2mm}, \hspace{0.3mm} 3})
\end{array}\right. & \hspace{-0.7mm} \begin{array}{l} \hspace{6mm}
\mbox{for} \hspace{3.4mm} j=1 \\ \hspace{6mm} \mbox{for}
\hspace{3.4mm} j=2 \\ \hspace{6mm} \mbox{for} \hspace{3.4mm} j=3 \: .
\end{array}
\end{array}
\end{equation}
The variable $\mu_j$ keeps track of whether each of the three
vertical lines in the double-box graph in
figure~\ref{fig:general_double_box} is part of some three-point
vertex or not, and respectively equals zero or one. For
mnemonic convenience, we will denote the values of $\mu_j$ by letters
as follows
\begin{equation}
\begin{array}{rll}
\mu_j = \mbox{m} \hspace{5mm} &\Longleftrightarrow& \hspace{5mm} \mu_j = 0 \\
\mu_j = \mbox{M} \hspace{5mm} &\Longleftrightarrow& \hspace{5mm} \mu_j = 1 \: .
\end{array}
\end{equation}
Finally, the notation $I_{(N_1, N_2, N_3, N_4, N_5, N_6)}$ will be used to
refer to a double-box integral with $N_i$ external massless legs
attached to vertex $i$.
\\
\\
To give an example of how three-point vertices play a role in
determining the number of kinematical solutions, let us consider
the third equation in eq.~(\ref{eq:onshell-values_alpha_beta}),
\begin{equation}
\alpha_3\alpha_4 \hspace{0.5mm}\propto\hspace{0.5mm} S_1 S_2 \: .
\end{equation}
We observe that if the right hand side is nonzero, one gets an
invertible relation between $\alpha_3$ and $\alpha_4$, leaving
either of them as an equivalent free parameter. If, on the other
hand, the right hand side is zero, this equation has two distinct
solutions, $\alpha_3=0$ or $\alpha_4=0$. The latter situation
occurs whenever the leftmost vertical line of the double-box graph is
part of some three-point vertex (in the above notation denoted by
$m$), and the splitting of one into two solutions of
eq.~(\ref{eq:onshell-values_alpha_beta}) is a reflection of the
existence of two types of massless on-shell three-point vertices
in $3+1$ dimensions \cite{Witten:2003nn}.
Indeed, assuming for the moment that
$S_1 = 0$, it follows from eq.~(\ref{eq:l1_parametrization}) that
$\alpha_3=0$ implies
\begin{equation}
 |\ell_1] \hspace{0.5mm}\propto\hspace{0.5mm} |\ell_1 - K_1] \hspace{0.5mm}\propto\hspace{0.5mm} |K_1] \label{aligned1}
\end{equation}
whereas $\alpha_4 = 0$ implies
\begin{equation}
 \langle \ell_1| \hspace{0.5mm}\propto\hspace{0.5mm}  \langle \ell_1 - K_1|
 \hspace{0.5mm}\propto\hspace{0.5mm}  \langle K_1| \: . \label{aligned2}
\end{equation}
Thus we see that the three-point vertex is special in that it
connects momenta with the property that either, as in eq.
(\ref{aligned1}), their square-bracket spinors are aligned, or, as
in eq. (\ref{aligned2}), their angle-bracket spinors are aligned.
In the following, we shall respectively denote the two cases in
eqs. (\ref{aligned1}) and  (\ref{aligned2}) with a $\ominus$ and
$\oplus$ label, and refer to a label as the chirality of the
vertex. Unlike a three-point vertex, a vertex with four or more
legs does not have a well-defined chirality.

Below we discuss the number of kinematical solutions to the
heptacut constraints (\ref{eq:on-shell_constraint_1})-(\ref{eq:on-shell_constraint_7}) and the intersections of their associated
Riemann surfaces for each of a total of four cases. These four
cases are defined by having all, exactly two, exactly one and none
of the vertical lines in the double-box graph be part of some
three-point vertex.

\subsection{Case 1: $(\mu_1,\mu_2,\mu_3)$ = (m,\hspace{0.7mm}m,\hspace{0.7mm}m)}\label{sec:Case_1}

Let us first consider the integral topologies where each vertical
line of the double-box graph is part of at least one three-point vertex.
This category includes, for instance, all topologies with four or
five massless external states, but also an infinite sequence of
topologies at higher points.

The case of four external massless states was studied in detail in
ref.~\cite{Kosower:2011ty} where it was shown that the number of
kinematical solutions to the heptacut constraints (\ref{eq:on-shell_constraint_1})-(\ref{eq:on-shell_constraint_7}) is six -- as
explained there, the solutions are uniquely characterized by the
distribution of chiralities ($\oplus$ or $\ominus$) at the
three-point vertices. Disallowed distributions are those with
uninterrupted chains of same-chirality vertices along the vertical
or horizontal lines; all other configurations are allowed for
generic external kinematics.

\subsubsection{Classification of kinematical solutions}\label{sec:Case_1_classification_of_solutions}

As it turns out, the classification of kinematical solutions in
the case of four massless external states extends uniformly to
cover all case 1 topologies. We can establish this in two steps.

Let us start by noting that for an allowed solution, three-point
vertices lying on a common vertical line must have opposite
chirality. For the leftmost and rightmost lines, this is already
visible from the discussion around eqs.~(\ref{aligned1}) and
(\ref{aligned2}). For example, having two $\oplus$ vertices on the
leftmost vertical line cannot be achieved for generic external
momenta because it would require
\begin{equation}
 \langle K_1 | \hspace{0.5mm}\propto\hspace{0.5mm}  \langle \ell_1 - K_1|
 \hspace{0.5mm}\propto\hspace{0.5mm}  \langle K_2| \label{allprop}
\end{equation}
and thus $K_1{\cdot}K_2=0$. Similarly, for generic external
momenta, forcing two $\oplus$ vertices to appear in the
central vertical line can be shown to require the four-momentum in
the central propagator to vanish, leaving no free parameter.

This already places an upper bound of $2^3=8$ on the number of
kinematical solutions. Not all of these configurations are
allowed, however.  For instance, three $\oplus$ vertices lying on
a horizontal line would force all angle bracket spinors on that
line to be proportional to each other, in analogy with
eq.~(\ref{allprop}).

\begin{figure}[!h]
\begin{center}
\includegraphics[angle=0, width=0.47\textwidth]{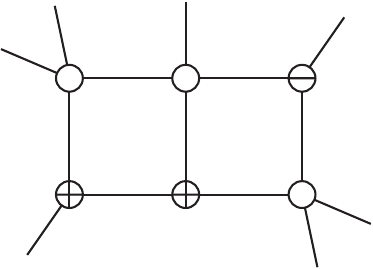}
\caption{A generic integral belonging to case 1: all three
vertical lines are part of some three-point vertex. The shown
chirality assignment is forbidden, as explained in the main
text.}\label{fig:Case_1_generic}
\end{center}
\end{figure}

More generally, consider the distribution of chiralities at the
vertices of the 7-particle topology shown in
figure~\ref{fig:Case_1_generic}.

\noindent This distribution of chiralities does not obviously
impose any constraints on the external kinematics. Closer
inspection, however, reveals that this assignment contains a
one-loop sub-box on the right loop (of the two-mass-easy type),
with opposite chiralities at its two massless corners.
But as is known from studies of one-loop boxes,
for generic external momenta, opposite corners of a
two-mass-easy box must have identical chiralities \cite{Britto:2004nc}. Thus,
the configuration in figure~\ref{fig:Case_1_generic} in fact does impose constraints on the external
momenta and is therefore disallowed. We have verified, by
exhaustion, that all chirality assignments not forbidden in such
ways lead to healthy kinematical solutions.

To summarize this discussion, we have derived three simple rules
which establish that there are exactly six kinematical solutions
for all topologies within case 1:

\begin{itemize}
\item Rule 1. Two same-chirality vertices cannot appear in a
vertical line. \item Rule 2. Three same-chirality vertices cannot
appear in an horizontal line. \item Rule 3. In rule 2,
opposite-chirality vertices at opposite corners of a one-loop
sub-box should be counted like same-chirality adjacent vertices
(cf. the right loop in figure \ref{fig:Case_1_generic}).
\end{itemize}

\subsubsection{Interpretation of Jacobian poles}

\begin{figure}[!h]
\begin{center}
\includegraphics[angle=0, width=0.53\textwidth]{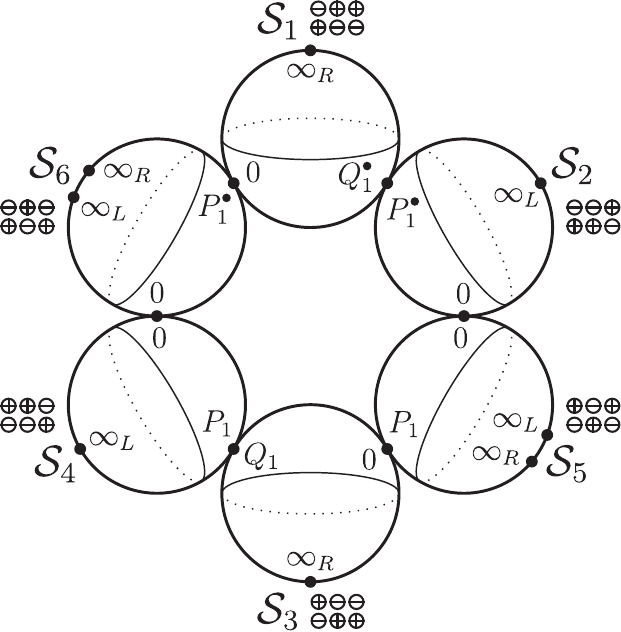}
\caption{The six different classes of kinematical solutions to the
heptacut constraints (\ref{eq:on-shell_constraint_1})-(\ref{eq:on-shell_constraint_7}),
illustrated here as Riemann spheres
(intended to represent the complex degree of freedom $z$ left
unfrozen by the heptacut constraints), in case 1. The kinematical
solutions are characterized by the distribution of chiralities at
the vertices of the double-box graph (see figure
\ref{fig:general_double_box}), shown next to each sphere. Each
Riemann sphere coincides with the two adjacent spheres in the
chain at a single point, illustrated as a black dot. These points
are precisely the poles of the heptacut Jacobian. The Riemann
spheres contain additional singularities, denoted by $\infty_L$ and
$\infty_R$, associated with respectively the left or right loop
momentum becoming infinite (respectively occuring as $z=\infty$
or $z=P_2^{(\bullet)}$ in
eqs.~(\ref{eq:Jac_poles_spinor_ratios})-(\ref{eq:parametrization_of_six_solutions})).
Parity-conjugate kinematical solutions
appear antipodally in the chain.}\label{fig:Jacobian_poles_1}
\end{center}
\end{figure}

As it turns out, the six kinematical solutions to the heptacut
constraints (\ref{eq:on-shell_constraint_1})-(\ref{eq:on-shell_constraint_7}) are not completely disjoint: as illustrated in
figure~\ref{fig:Jacobian_poles_1}, for any given kinematical
solution $\mathcal{S}_i$, depicted there as a Riemann sphere,
there are two special points where it coincides with a different
kinematical solution.  We now proceed to locate these special
points and show that the six Riemann spheres link into a chain.

\def\SS{\mathcal{S}}

To clarify the exposition, we will consider a particular
representative of case 1 and write out explicitly the kinematical
solutions and the Jacobian determinants. For example, let us
choose the integral topology whose vertex momenta (as defined
in figure \ref{fig:general_double_box}) are
\begin{eqnarray}
K_1 &=& k_1 + k_2 \: ,
\hspace{1cm} K_2 = k_3 \: , \hspace{10mm} K_3 = 0 \: , \\
K_4 &=& k_4 \: , \hspace{1.84cm} K_5 = k_5 \: , \hspace{10mm} K_6
= 0 \: ,
\end{eqnarray}
with the $k_i$ being lightlike vectors. In terms of the spinor ratios
\begin{equation}
\begin{array}{lll} P_1 = -\frac{\langle K_1^\flat \hspace{0.4mm} k_5 \rangle}
{2\langle k_3 \hspace{0.4mm} k_5 \rangle} \: , \hspace{10mm} & P_2
= -\frac{\langle K_1^\flat \hspace{0.4mm} k_4 \rangle}{2\langle
k_3 \hspace{0.4mm} k_4 \rangle} \: , \hspace{10mm} &
Q_1 = -\frac{[k_5 \hspace{0.4mm} K_1^\flat]}{2[k_4 \hspace{0.4mm} K_1^\flat]} \: , \\[2mm]
P_1^\bullet = -\frac{[K_1^\flat \hspace{0.4mm} k_5]} {2[k_3
\hspace{0.4mm} k_5]} \: , \hspace{10mm} & P_2^\bullet =
-\frac{[K_1^\flat \hspace{0.4mm} k_4]}{2[k_3 \hspace{0.4mm} k_4]}
\: , \hspace{10mm} & Q_1^\bullet = -\frac{\langle k_5
\hspace{0.4mm} K_1^\flat \rangle}{2\langle k_4 \hspace{0.4mm}
K_1^\flat \rangle} \: ,
\end{array} \label{eq:Jac_poles_spinor_ratios}
\end{equation}
the six kinematical solutions $\mathcal{S}_1, \ldots,
\mathcal{S}_6$ are obtained by fixing the parameters of the loop
momenta to the values
\begin{equation}
\mbox{$\mathcal{S}_1, \ldots, \mathcal{S}_6$:} \hspace{3mm}
\left\{ \hspace{-0.5mm}
\begin{array}{ll} \alpha_1 = 1 \: , &
\hspace{7mm} \beta_1 = 0 \\
\alpha_2 = 0 \: , & \hspace{7mm} \beta_2 = 1
\end{array} \right.
\end{equation}
\begin{equation}
\begin{array}{ll} \mbox{$\mathcal{S}_1$:} \hspace{3mm} \left\{ \hspace{-0.5mm}
\begin{array}{ll} \alpha_3 = P_1^\bullet \: , & \hspace{4mm} \beta_3 = z \\
\alpha_4 = 0 \: , & \hspace{4mm} \beta_4 = 0 \end{array}
\hspace{1mm}; \right.
& \hspace{4mm} \mbox{$\mathcal{S}_2$:} \hspace{3mm} \left\{ \hspace{-0.5mm} \begin{array}{ll} \alpha_3 = z \: , & \hspace{4mm} \beta_3 = Q_1^\bullet \\
\alpha_4 = 0 \: , & \hspace{4mm} \beta_4 = 0 \end{array} \right. \\[6mm]
\mbox{$\mathcal{S}_3$:} \hspace{3mm} \left\{ \hspace{-0.5mm} \begin{array}{ll} \alpha_3 = 0 \: , & \hspace{4mm} \beta_3 = 0 \\
\alpha_4 = P_1 \: , & \hspace{4mm} \beta_4 = z \end{array}
\hspace{1mm} ;
\right. & \hspace{4mm} \mbox{$\mathcal{S}_4$:} \hspace{3mm} \left\{ \hspace{-0.5mm} \begin{array}{ll} \alpha_3 = 0 \: , & \hspace{4mm} \beta_3 = 0 \\
\alpha_4 = z \: , & \hspace{4mm} \beta_4 = Q_1 \end{array} \right. \\[6mm]
\mbox{$\mathcal{S}_5$:} \hspace{3mm} \left\{ \hspace{-0.5mm}
\begin{array}{ll} \alpha_3 = 0 \: , & \hspace{4mm} \beta_3 =
-\frac{\langle k_5 \hspace{0.4mm} k_3
\rangle (z-P_1)}{2\langle k_4 \hspace{0.4mm} k_3 \rangle (z-P_2)} \\
\alpha_4 = z \: , & \hspace{4mm} \beta_4 = 0 \end{array}
\hspace{0.5mm} ;
\right. & \hspace{4mm} \mbox{$\mathcal{S}_6$:} \hspace{3mm} \left\{ \hspace{-0.5mm} \begin{array}{ll} \alpha_3 = z \: , & \hspace{4mm} \beta_3 = 0 \\
\alpha_4 = 0 \: , & \hspace{4mm} \beta_4 = -\frac{[k_5
\hspace{0.4mm} k_3] (z-P_1^\bullet)}{2[k_4 \hspace{0.4mm} k_3]
(z-P_2^\bullet)} \end{array} \right.
\end{array}\label{eq:parametrization_of_six_solutions}
\end{equation}
with $z \in \mathbb{C}$ a free parameter. The associated heptacut Jacobians are
\begin{equation} \hspace{-3mm}
J_i(z) %\hspace{1mm}
=
%\hspace{1mm}
%\frac{1}{(s_{13}+s_{23}) s_{45} 2K_1^\flat {\cdot}k_5}
\frac{1}{s_{45}\big ((s_{13}+s_{23})(s_{15}+s_{25})-s_{12}s_{35})\big)}
\!\times\! \left\{
\begin{array}{ll}
\pm \hspace{0.4mm} \big(  \hspace{0.2mm} z (1-z/P_1^\bullet) \hspace{0.3mm}\big)^{-1} & \hspace{4mm} \mathrm{for} \hspace{3mm} i=2,6 \\[1mm]
\pm \hspace{0.4mm} \big( \hspace{0.2mm} z (1-z/P_1) \hspace{0.3mm}\big)^{-1} & \hspace{4mm} \mathrm{for} \hspace{3mm} i=4,5 \\[1mm]
- \hspace{0.4mm} \big( \hspace{0.2mm} z (1-z/Q_1^\bullet) \hspace{0.3mm}\big)^{-1} & \hspace{4mm} \mathrm{for} \hspace{3mm} i=1 \\[1mm]
- \hspace{0.4mm} \big( \hspace{0.2mm} z (1-z/ Q_1) \hspace{0.3mm}\big)^{-1} & \hspace{4mm}
\mathrm{for} \hspace{3mm} i=3 \: .
\end{array} \right. \label{eq:Jacobians_Case_1}
\end{equation}
In the first two lines of eq.~(\ref{eq:Jacobians_Case_1}),
the plus or minus signs refer respectively to the first or second indicated kinematical solution.
Now, consider first the
intersection between $\SS_4$ and $\SS_6$. In $\SS_4$ we have
$\alpha_3=0$ but $\alpha_4$ free, while in $\SS_6$ we have
$\alpha_4=0$ and $\alpha_3$ free; the intersection is simply a
point, located in $\mathcal{S}_4$ at $\alpha_4=0$ and in
$\mathcal{S}_6$ at $\alpha_3=0$. At this point, $\beta_3$
equals zero while $\beta_4$ takes on a finite value explicitly given in
eqs.~(\ref{eq:Jac_poles_spinor_ratios}) and (\ref{eq:parametrization_of_six_solutions}).

To understand better why $\mathcal{S}_4$ and $\mathcal{S}_6$
coincide at a point, let us examine what is happening to the loop
momentum $\ell_1$ at $z=0$ in $\mathcal{S}_4$. By assumption,
either vertex 1 or 2 in figure~\ref{fig:general_double_box} is a three-point vertex; let us consider here
the former case.   It is straightforward to see that in the
parametrization (\ref{eq:l1_parametrization}) the on-shell
constraints (\ref{eq:on-shell_constraint_1})-(\ref{eq:on-shell_constraint_3}) and (\ref{eq:on-shell_constraint_7})
are solved within $\SS_6$ by setting
$(\alpha_1,\alpha_2,\alpha_3,\alpha_4)=(
\frac{S_2+\gamma_1}{\gamma_1},0,z,0)$. At $z=0$ we then observe
that
\begin{equation}
 \ell_1^\mu \hspace{0.5mm}=\hspace{0.5mm} \frac{S_2+\gamma_1}{\gamma_1} K_1^\mu
 \hspace{1mm} \propto \hspace{1mm} K_1^\mu \: ,  \label{soll1}
\end{equation}
i.e., the loop momentum is collinear with that of a massless external
particle.

This collinearity can be immediately understood from
figure~\ref{fig:Jacobian_poles_1}: at the intersection of $\SS_4$
and $\SS_6$, the lower-left vertex must simultaneously be of the
$\ominus$ and $\oplus$ type, and therefore the momenta connected
by this vertex are mutually collinear. Moreover, when both of the momenta $K_1$
and $K_2$ are massless, the simultaneous collinearity conditions
at the two left-most vertices imply that the momentum
of the particle exchanged between these vertices must vanish.
Indeed, we see that in this case, eq.~(\ref{soll1}) implies that
$\ell_1-K_1=0$. Physically, this corresponds to a soft divergence
region, giving rise to an infrared divergence in the original
two-loop integral. In a gauge theory, the exchanged soft particle
producing such a singularity will necessarily be a soft gluon,
as can be argued from the behavior of the three-point vertices.

Similarly, the intersection between $\SS_2$ and $\SS_5$ occurs at
a point where $\alpha_3=\alpha_4=0$. By symmetry, there are
similar intersections at points where $\beta_3=\beta_4=0$, merging
$\SS_1$ with $\SS_6$, and $\SS_3$ with $\SS_5$. Finally, the
intersections between $\SS_1$ and $\SS_2$, and between $\SS_3$ and
$\SS_4$, occur at points where the momenta in the central
three-point vertices become collinear.
\\
\\
Let us conclude this discussion by the observation that the poles of the
Jacobian determinants in eq. (\ref{eq:Jacobians_Case_1}) coincide
with the intersection points of the six Riemann spheres shown in
figure \ref{fig:Jacobian_poles_1}. We have checked that this
phenomenon extends to all integral topologies in case 1.

\subsubsection{Residue relations across solutions}

At the location of any Jacobian pole, each loop momentum $\ell_i$
as evaluated from either of two intersecting spheres assumes
identical values; for example,
\begin{equation}
\ell_i (0) \big|_{\mathcal{S}_4} \hspace{1mm}=\hspace{1mm} \ell_i
(0) \big|_{\mathcal{S}_6} \hspace{6mm} \mbox{for} \hspace{3mm}
i=1,2 \: .
\end{equation}
As a result of this, given an arbitrary function $f(\ell_1 (z),
\ell_2 (z))$ that does not share these poles, one has the
identities
\begin{eqnarray}
\mathop{\mathrm{Res}}_{z=Q_1^\bullet} J(z) \hspace{0.4mm} f(\ell_1
(z), \ell_2 (z) ) \big|_{\mathcal{S}_1}
\hspace{1mm}&=&\hspace{1mm} -\mathop{\mathrm{Res}}_{z=P_1^\bullet}
J(z) \hspace{0.4mm}
f(\ell_1 (z), \ell_2 (z) ) \big|_{\mathcal{S}_2} \label{eq:residues_of_cuts_relation_1}\\
\mathop{\mathrm{Res}}_{z=0} J(z) \hspace{0.4mm} f(\ell_1 (z),
\ell_2 (z) ) \big|_{\mathcal{S}_2} \hspace{1mm}&=&\hspace{1mm}
-\mathop{\mathrm{Res}}_{z=0}
J(z) \hspace{0.4mm} f(\ell_1 (z), \ell_2 (z) ) \big|_{\mathcal{S}_5} \label{eq:residues_of_cuts_relation_2}\\
\mathop{\mathrm{Res}}_{z=P_1} J(z) \hspace{0.4mm} f(\ell_1 (z),
\ell_2 (z) ) \big|_{\mathcal{S}_5} \hspace{1mm}&=&\hspace{1mm}
-\mathop{\mathrm{Res}}_{z=0} J(z) \hspace{0.4mm} f(\ell_1 (z),
\ell_2 (z) ) \big|_{\mathcal{S}_3} \label{eq:residues_of_cuts_relation_3}\\
\mathop{\mathrm{Res}}_{z=Q_1} J(z) \hspace{0.4mm} f(\ell_1 (z),
\ell_2 (z) ) \big|_{\mathcal{S}_3} \hspace{1mm}&=&\hspace{1mm}
-\mathop{\mathrm{Res}}_{z=P_1} J(z) \hspace{0.4mm} f(\ell_1 (z),
\ell_2 (z) ) \big|_{\mathcal{S}_4} \label{eq:residues_of_cuts_relation_4}\\
\mathop{\mathrm{Res}}_{z=0} J(z) \hspace{0.4mm} f(\ell_1 (z),
\ell_2 (z) ) \big|_{\mathcal{S}_4} \hspace{1mm}&=&\hspace{1mm}
-\mathop{\mathrm{Res}}_{z=0} J(z) \hspace{0.4mm} f(\ell_1 (z),
\ell_2 (z) ) \big|_{\mathcal{S}_6} \label{eq:residues_of_cuts_relation_5}\\
\mathop{\mathrm{Res}}_{z=P_1^\bullet} J(z) \hspace{0.4mm} f(\ell_1
(z), \ell_2 (z) ) \big|_{\mathcal{S}_6}
\hspace{1mm}&=&\hspace{1mm} -\mathop{\mathrm{Res}}_{z=0} J(z)
\hspace{0.4mm} f(\ell_1 (z), \ell_2 (z) ) \big|_{\mathcal{S}_1} \:
. \label{eq:residues_of_cuts_relation_6}
\end{eqnarray}
We note that the uniform pattern of signs owes to the conventions explained below
eq.~(\ref{hexacut}). As explained in appendix \ref{sec:cuts_across_kin_sols}, these
identities can be applied in computations of heptacut two-loop
amplitudes $\left. J(z) \hspace{0.2mm} \prod_{j=1}^6
A_j^\mathrm{tree}(z) \right|_{\mathcal{S}_i}$ to explain the
vanishing of certain residues, as well as the seemingly accidental
equality between pairs of other residues. In generalized-unitarity
calculations, such residues form the input out of which the
integral coefficients of the two-loop amplitude are computed.

More importantly, as we will explain in section
\ref{sec:proliferation_of_contours}, there are equivalence relations
dual to the identities
(\ref{eq:residues_of_cuts_relation_1})-(\ref{eq:residues_of_cuts_relation_6})
which explain the proliferation of the heptacut contours of ref.
\cite{Kosower:2011ty} as a simple redundancy of variables.
\\
\\
To summarize section~\ref{sec:Case_1}, we have found that in case 1 there are six classes of kinematical
solutions to the heptacut constraints (\ref{eq:on-shell_constraint_1})-(\ref{eq:on-shell_constraint_7}), each of which is labeled by
a free complex variable $z\in \mathbb{C}$, parametrizing
a Riemann sphere. The six Riemann spheres thus associated with the
kinematical solutions intersect pairwise in six points, linking
into a chain as illustrated in figure~\ref{fig:Jacobian_poles_1}.
Within each sphere, the Jacobian factor that arose from
linearizing the cut constraints gives rise to a measure which has
two poles, located at the intersection with the neighboring
spheres in the chain. These poles were called hidden or composite leading
singularities in refs.~\cite{Buchbinder:2005wp,Cachazo:2008dx,Cachazo:2008vp}. Pleasingly, we find that they are
directly related to the physical collinear and infrared
singularities of the theory.\footnote{One may inquire about the
Jacobian poles coming from three-point vertices in the center of
the double-box graph. While these certainly do correspond to a dangerous
infrared-singular region of integration, in this case there is not
necessarily a divergence. For instance, the fully massive
four-point double box, belonging to case 3 below, is infrared
finite.}

\subsection{Case 2: $(\mu_1,\mu_2,\mu_3)$ = (M,\hspace{0.7mm}m,\hspace{0.7mm}m),
(m,\hspace{0.7mm}M,\hspace{0.7mm}m) or (m,\hspace{0.7mm}m,\hspace{0.7mm}M)}\label{sec:Case_2}

\begin{figure}[!h]
\begin{center}
\includegraphics[angle=0, width=0.98\textwidth]{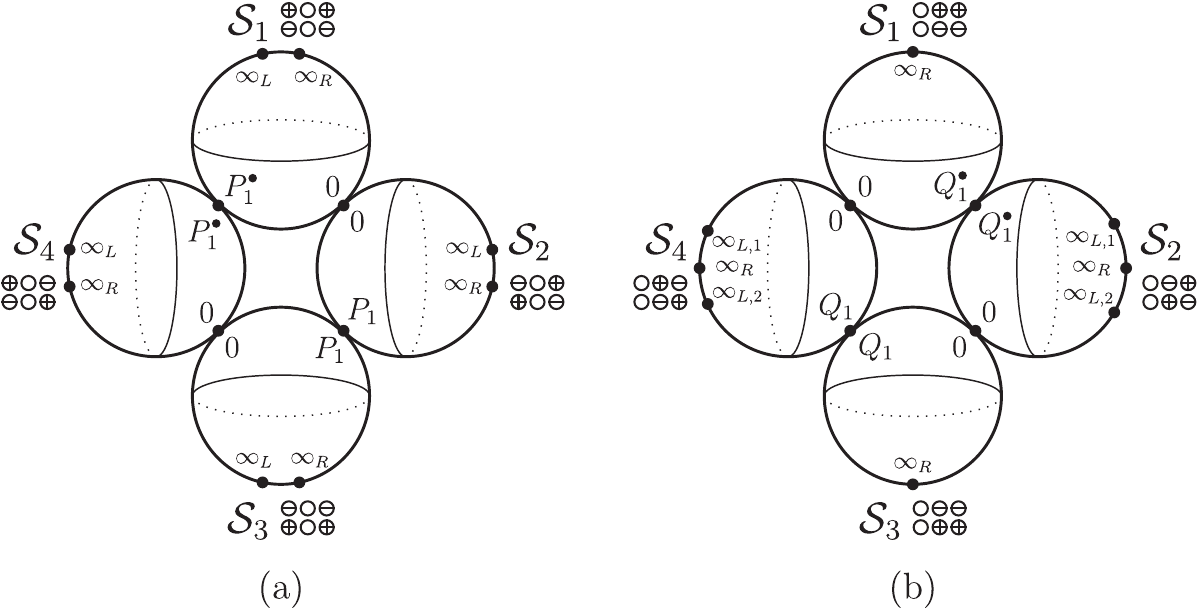}
\caption{The four different classes of kinematical solutions to
the heptacut constraints in case 2: exactly two vertical lines of
the double-box graph are part of some three-point vertex. Figure
(a) illustrates the (m, \hspace{-1.5mm} M, \hspace{-1.5mm} m)
subcase, whereas figure (b) illustrates the (M, \hspace{-1.5mm} m,
\hspace{-1.5mm} m) subcase. The only difference between these two subcases
is the number of points contained in each sphere at which one of
the loop momenta becomes infinite. We observe that in both subcases the
number of independent residues one can take is
8.}\label{fig:Jacobian_poles_2}
\end{center}
\end{figure}

\noindent This is the case where exactly two vertical lines of the double-box graph
are part of some three-point vertex.  The analysis leading to
rules 1-3 in section \ref{sec:Case_1_classification_of_solutions}
remains valid and shows that there are exactly four kinematical
solutions for any topology belonging to this case. These solutions
are uniquely characterized by those assignments of vertex
chiralities where no two $\oplus$ or $\ominus$ occur in the same
vertical line. Based on our insights from the previous subsection,
it is natural to expect that the four Riemann spheres associated
with the kinematical solutions again have pointwise intersections
and are linked into a chain. This picture turns out to be correct,
and we will now elaborate on some of its details.

For the double-box topologies of the type
(m,\hspace{0.7mm}M,\hspace{0.7mm}m), one finds that the
intersection points coincide with the poles of the Jacobian in
eq.~(\ref{eq:heptacut_Jacobian_S_2_and_S_6}), in complete analogy
with case~1. Moreover, each of these intersection points is
associated with the simultaneous collinearity of the momenta in
some three-point vertex of the double-box graph, again in exact analogy
with case 1. The kinematical solutions of subcase (m,\hspace{0.7mm}M,\hspace{0.7mm}m)
are illustrated in figure~\ref{fig:Jacobian_poles_2}(a).

%In order to characterize the intersection points, let us consider
%the form of the heptacut Jacobian. As above, we expect it to have
%poles coinciding with these points. Let us first consider the
%topologies of the type (m,\hspace{0.7mm}M,\hspace{0.7mm}m), in
%figure~\ref{fig:Jacobian_poles_2}a. All four spheres are
%qualitatively similar, and so we consider only $\SS_\#$ where
%$\alpha_4=\beta_4=0$. Solving for $\beta_3$ as a function of
%$z\equiv \alpha_3$, the Jacobian in
%eq.~(\ref{eq:heptacut_Jacobian_S_2_and_S_6}) gives
%\begin{equation}
%J = \oint_\Gamma \frac{dz}{z \hspace{0.4mm} B_0(z)}
%\end{equation}
%where we have used that $B_{-1}=0$; $B_0(z)$ is a linear function
%of $z$. On the $z=0$ pole, the solution merges with $\SS_\#$,
%while on the $B_0(z)=0$ pole, $\beta_3=0$ and the solution merges
%with $\SS_\#$.

Seemingly, a new technical issue arises for the topologies of type
(M,\hspace{0.7mm}m,\hspace{0.7mm}m):\footnote{By assumption, at
least one of the vertex momenta $K_3$ or $K_6$ (defined in figure
\ref{fig:general_double_box}) vanishes. Without loss of
generality, we take $K_6$ to vanish here.} the square root in eq.
(\ref{eq:heptacut_Jacobian_S_2_and_S_6}) suggests that the
Jacobian contains branch cuts.  Despite appearances, the radicand
is in fact a perfect square, as can easily be seen by writing the
central-propagator condition $0 = (\ell_1 + \ell_2)^2=\langle
\ell_1 \hspace{0.5mm} \ell_2\rangle [\ell_1 \hspace{0.5mm}
\ell_2]$ in the factorized form
\begin{eqnarray}
(\ell_1 + \ell_2)^2 &=& \frac{1}{\alpha_1^2\beta_2^2} \left(
\big(\alpha_1[ K_1^{\flat}| + \alpha_3 [K_2^{\flat}|
\big)\big(\beta_2 | K_5^\flat] + \beta_4 |K_4^\flat] \big)\right) \nonumber \\
&\phantom{=}& \hspace{20mm} \times \left( \big (\alpha_1\langle
K_1^{\flat}| + \alpha_4 \langle K_2^{\flat}| \big) \big(\beta_2 |
K_5^\flat\rangle + \beta_3 |K_4^\flat\rangle \big)\right) \nonumber \\
&=& 0 \: . \label{Mmm_central}
\end{eqnarray}
Plugging eq.~(\ref{Mmm_central}) into eq.~(\ref{hexacut}) and performing
the last integration yields an explicitly rational formula
for the Jacobian, containing poles rather than branch cuts.

The (M,\hspace{0.7mm}m,\hspace{0.7mm}m) subcase thus presents no new features compared to the
(m,\hspace{0.7mm}M,\hspace{0.7mm}m) subcase, except in one regard, illustrated in
figure~\ref{fig:Jacobian_poles_2}(b): the number of points in each Riemann
sphere at which one of the loop momenta becomes infinite. This can
be understood as follows. For solutions $\mathcal{S}_2$ and $\mathcal{S}_4$,
the fact that $S_1 S_2 \neq 0$ implies that $\alpha_3 \sim z$ and
$\alpha_4 \sim \frac{1}{z}$, allowing for two distinct points on each
of these spheres at which the left loop momentum $\ell_1$ becomes infinite, denoted
respectively as $\infty_{L,1}$ and $\infty_{L,2}$ in figure~\ref{fig:Jacobian_poles_2}(b).
On the other hand, in solutions $\mathcal{S}_1$ and $\mathcal{S}_3$ the loop
momentum $\ell_1$ assumes a constant value independent of $z$.\footnote{To be
somewhat more detailed, the chirality distributions in
solutions $\mathcal{S}_1$ and $\mathcal{S}_3$ allow us to construct
an ``effective'' one-loop box, obtained by collapsing the horizontal
propagators between same-chirality labels on the right loop. As it turns
out, the solution to the quadruple-cut constraints of this one-loop
box is exactly equal to the left loop momentum $\ell_1$ in the original
heptacut double box. But as the quadruple cut freezes all components of the
one-loop box momentum, this implies that the $\ell_1$ obtained from
$\mathcal{S}_1$ and $\mathcal{S}_3$ cannot have any dependence on $z$.}
In particular, neither of these spheres contain any point at which
$\ell_1$ becomes infinite.

The above reasoning readily extends to case 1 and explains the positioning
of the infinity poles in figure~\ref{fig:Jacobian_poles_1}.
\\
\\
From figures~\ref{fig:Jacobian_poles_1} and
\ref{fig:Jacobian_poles_2} we observe that in both case 1 and 2, the
number of independent residues one can take is 8. Here, the qualifier ``independent''
refers to the fact that on any given Riemann sphere, the residues necessarily
add to zero, allowing any one residue to be expressed in terms
of the remaining ones on the sphere. Thus, counting the number of poles shown in
figures~\ref{fig:Jacobian_poles_1} and \ref{fig:Jacobian_poles_2},
and subtracting the number of Riemann spheres to compensate
for the redundancy, we find 8 independent residues in all cases
considered so far.

%Solutions $\SS_2$ and $\SS_3$ are then obtained by setting the
%factor in the first line of eq. (\ref{Mmm_central}) to zero, and
%the other two are obtained by setting the factor in the second
%line to zero. We note that in $\SS_3$, in which the first factor
%is set to zero and $\beta_4=0$, $\ell_1$ is non-dynamical: the
%values of all $\alpha_i$ parameters are fixed independent of the
%free variable $z\equiv\beta_3$.  On the other hand, in $\SS_2$ we
%have $\beta_3=0$ and $\ell_1$ contains both $\sim z$ and a $\sim
%\frac1z$ dependence on $z\equiv\beta_4$.

\subsection{Case 3: $(\mu_1,\mu_2,\mu_3)$ = (M,\hspace{0.7mm}M,\hspace{0.7mm}m),
(M,\hspace{0.7mm}m,\hspace{0.7mm}M) or (m,\hspace{0.7mm}M,\hspace{0.7mm}M)}\label{sec:Case_3}

\begin{figure}[!h]
\begin{center}
\includegraphics[angle=0, width=0.56\textwidth]{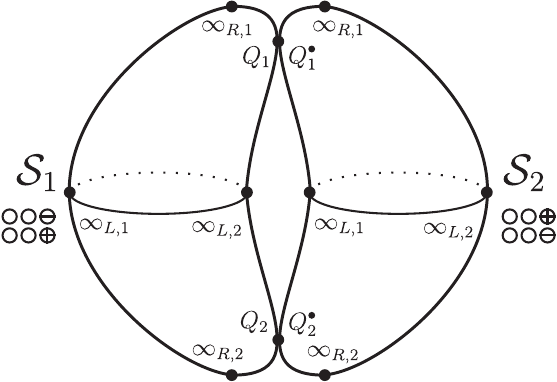}
\caption{The two different classes of kinematical solutions to the
heptacut constraints in subcase
(M,\hspace{0.7mm}M,\hspace{0.7mm}m) of case 3. The subcases
(M,\hspace{0.7mm}m,\hspace{0.7mm}M) and
(m,\hspace{0.7mm}M,\hspace{0.7mm}M) are
similar. Again the
number of independent residues one can take is
8.}\label{fig:Jacobian_poles_3}
\end{center}
\end{figure}

In double-box topologies where exactly one vertical line of the
graph is part of some three-point vertex, the rules of section
\ref{sec:Case_1_classification_of_solutions} imply that there are
two kinematical solutions.

In the (M,\hspace{0.7mm}M,\hspace{0.7mm}m) case, one of the two
Riemann spheres can be parametrized by $z\equiv \alpha_3$ (and its
parity conjugate by $z\equiv \alpha_4$), and so the equations
(\ref{eq:heptacut_Jacobian_S_2_and_S_6})-(\ref{eq:coeff_B_-1})
readily apply. Because $B_{-1}=0$, the Jacobian
(\ref{eq:heptacut_Jacobian_S_2_and_S_6}) is manifestly a rational
function and has only poles in $z$. Exactly as in the previous
cases, these poles are located at the intersections of the Riemann
spheres; in particular, the Jacobian has exactly two poles on each
sphere.

In the (M,\hspace{0.7mm}m,\hspace{0.7mm}M) case, assuming again
(without loss of generality) that $K_6=0$, we can proceed as in
the (M,\hspace{0.7mm}m,\hspace{0.7mm}m) case above, following
eq.~(\ref{Mmm_central}). The same expression remains valid here
and makes manifest the fact that the Jacobians are rational
functions (of the variables $\alpha_3, \alpha_4, \beta_3$ or
$\beta_4$). Again, the poles of the Jacobians coincide with the
intersections of the Riemann spheres corresponding to the
kinematical solutions.

The two kinematical solutions associated with the subcase
(M,\hspace{0.7mm}M,\hspace{0.7mm}m) are illustrated in figure~\ref{fig:Jacobian_poles_3}
which also shows that the number of independent residues one can take
is again 8, as in the previous cases. The other subcases
(M,\hspace{0.7mm}m,\hspace{0.7mm}M) and
(m,\hspace{0.7mm}M,\hspace{0.7mm}M) are
similar.

\subsection{Case 4: $(\mu_1,\mu_2,\mu_3)$ = (M,\hspace{0.7mm}M,\hspace{0.7mm}M)}\label{sec:Case_4}

\begin{figure}[!h]
\begin{center}
\includegraphics[angle=0, width=0.4\textwidth]{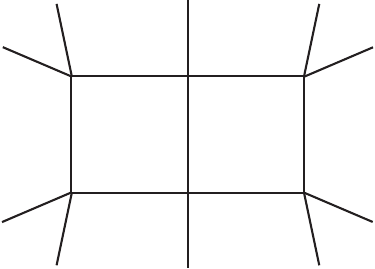}
\caption{The integral $I_{(2,2,1,2,2,1)}$, the simplest example of
an integral with more than three particles at all vertices, and
whose heptacut Jacobian $J(z)$ accordingly has branch cuts that
cannot be removed by any reparametrization $z \to \varphi(z)$. As
argued in the main text, this is presumably related to the appearance
of functions in the analytic expression for $I_{(2,2,1,2,2,1)}$
which cannot be expressed in terms of generalized
polylogarithms.\label{fig:elliptic10}}
\end{center}
\end{figure}

This is the case in which the double-box graph contains no
three-point vertices. For the scattering of massless particles,
the first time this occurs is for 10 particles, as depicted in
figure~\ref{fig:elliptic10}.

\begin{figure}[!h]
\begin{center}
\includegraphics[angle=0, width=0.8\textwidth]{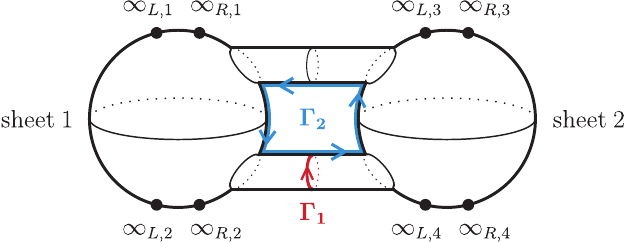}
\caption{(Color online). The single class of kinematical solutions
to the heptacut constraints in case 4 where the double-box graph
contains no three-point vertices. Here, the Jacobian develops branch
cuts, and the heptacut loop-momentum parameter $z$ thus parametrizes a two-sheeted
cover of the Riemann sphere.
The two sheets are shown glued together along their branch
cuts, illustrating how this Riemann surface is topologically
equivalent to an elliptic curve. The topological cycles
$\Gamma_1$ and $\Gamma_2$, respectively shown in red and blue,
enclose two distinct pairs of branch points and provide
natural contours of integration for the
heptacut double box in eq.~(\ref{ellipticint}). We observe that there are eight poles at which
one of the loop momenta becomes infinite.
In this case one finds a total of nine independent leading-singularity cycles, as explained in the main text.}\label{fig:elliptic_curve}
\end{center}
\end{figure}

To analyze this case, we return to
the Jacobian determinant in
eq.~(\ref{eq:heptacut_Jacobian_S_2_and_S_6}) which takes the form
\begin{equation}
 J=\oint_{\Gamma_i} \frac{dz}{\sqrt{Q(z)}} \label{ellipticint}
\end{equation}
where $Q(z)=z^2(B_0(z)^2-4B_1(z)B_{-1}(z))$ is a quartic
polynomial. Numerically, we find that for generic
10-particle kinematics, the four roots $r_i$ of this polynomial
are distinct. This means that, contrary to the previous cases,
the Jacobian contains genuine branch cuts (meaning that they cannot be removed
by any redefinition of $z$). The integration variable $z$ in eq.~(\ref{ellipticint})
therefore parametrizes a two-sheeted
cover of the Riemann sphere. This is topologically equivalent to an elliptic
curve (i.e., a genus one Riemann surface), as illustrated in figure~\ref{fig:elliptic_curve}.
In particular, there is a single class of kinematical solutions to the
heptacut constraints (\ref{eq:on-shell_constraint_1})-(\ref{eq:on-shell_constraint_7}) in this case.

For an elliptic curve there are two natural cycles over which the $z$ integration
in eq.~(\ref{ellipticint}) can be performed,
generalizing the notion of a residue
-- namely, its topological cycles $\Gamma_1$ and $\Gamma_2$,
respectively shown in red and blue in figure~\ref{fig:elliptic_curve}.  In
terms of the $z$ variable, these are cycles which enclose a pair
of branch points. Integrations over such cycles produce so-called
complete elliptic integrals of the first
kind $K(t)$ where the argument $t$ is some cross-ratio of the four roots of
the radicand $Q(z)$. As they arise when performing the loop
integration on a compact T${}^8$ contour, the integration cycles
$\Gamma_1$ and $\Gamma_2$ define leading
singularity cycles of the double-box integral.
\\
\\
As illustrated in figure~\ref{fig:elliptic_curve}, the number of poles
at which one of the loop momenta becomes infinite is 8. This can easily be explained as follows.
The fact that $S_1 S_2 \neq 0$ implies that $\alpha_3 \sim z$ and
$\alpha_4 \sim \frac{1}{z}$, allowing for two distinct points on each
sheet of the elliptic curve at which the left loop momentum $\ell_1$ becomes infinite;
these points are denoted as $\infty_{L,i}$ in the figure.
Moreover, since each of the sheets can equivalently be parametrized in terms
of $\beta_3$ or $\beta_4$, the fact that $S_4 S_5 \neq 0$ implies that $\beta_3 \sim z$ and
$\beta_4 \sim \frac{1}{z}$, allowing for two distinct points on each
sheet at which the right loop momentum $\ell_2$ becomes infinite;
these points are denoted as $\infty_{R,i}$. Thus, there are in total 8
poles on the elliptic curve.

The residues of these 8 poles are not all independent, however.  For instance, their sum is zero since it corresponds to a contractible cycle,
as can be seen from figure~\ref{fig:elliptic_curve}.  If all the infinity poles were only simple poles, which would be the case if all numerator
insertions in the double-box integral were at most linear
in each of the loop momenta $\ell_1$, $\ell_2$, there would exist a second, less obvious relation\footnote{This relation is easier to describe when the elliptic curve
is viewed as the complex plane modulo the doubly-periodic identification $z\simeq z+1, \hspace{0.8mm} z\simeq z+\tau$.
While the first relation mentioned in the main text arises from integrating a form $\omega(z)$ along the boundary of a fundamental domain,
the second relation arises from integrating $\omega(z) z$.  In the absence of double poles, the latter relation relates the sum of the residues weighted by $z$ to
integrals over $\Gamma_1$ and $\Gamma_2$. We refer the reader to ref.~\cite{milne2006} for more details;
in particular, to Chapter III, Proposition 2.1.}.
However, the large-momentum behavior of theories such as pure Yang-Mills or QCD are not such as to produce only simple poles,
and so this relation does not apply.
Including the two topological cycles $\Gamma_i$ in our counting, we thus find a total of 9 independent leading-singularity cycles in case 4, in contradistinction with the previous cases 1-3.
\\
\\
Finally, let us summarize our discussion of the number of
solutions to the maximal cut of the double-box graph in figure~\ref{fig:general_double_box}.
We have observed that as the number of three-point
vertices in the double-box graph grows, the number of associated
Riemann surfaces increases from one surface (of genus one) to two
spheres linked by pointwise intersections, and further on to four,
and finally six spheres thus linked. The branching of
kinematical solutions can be understood intuitively
from the observation that the equation $x_1 x_2 = m$ with
$m \neq 0$ has a single connected component
as its solution when $x_1$ and $x_2$ are allowed to be complex;
but two essentially-disconnected components when $m = 0$.
Applied to, e.g., the equation $\alpha_3 \alpha_4 \propto S_1 S_2$
in eq.~(\ref{eq:onshell-values_alpha_beta}), this insight leads us to
expect a splitting of one Riemann sphere
into two as $S_1 S_2 \to 0$.\footnote{We thank D. Kosower for this observation.}
This is indeed what happens, as
exemplified by the splitting of the sphere $\mathcal{S}_1$ in figure~\ref{fig:Jacobian_poles_3}
into the spheres $\mathcal{S}_3$ and $\mathcal{S}_4$ in figure~\ref{fig:Jacobian_poles_2}(a)
whose left-most pair of double-box vertices then acquires chiralities.
Taking the limit $\mu_2 \to \mathrm{m}$ of figure~\ref{fig:Jacobian_poles_2}(a),
the middle pair of vertices will acquire chiralities, splitting
the spheres $\mathcal{S}_2$ and $\mathcal{S}_4$ into the respective
pairs $(\mathcal{S}_1, \mathcal{S}_2)$ and $(\mathcal{S}_3, \mathcal{S}_4)$
in figure~\ref{fig:Jacobian_poles_1}. In contrast, the solutions $(\mathcal{S}_1, \mathcal{S}_3)$
in figure~\ref{fig:Jacobian_poles_2}(a) admit only one chirality
assignment to the middle vertices, as dictated by the
rules in section~\ref{sec:Case_1_classification_of_solutions},
and are transformed into the solutions $(\mathcal{S}_5, \mathcal{S}_6)$
in figure~\ref{fig:Jacobian_poles_1}. As the resulting six solutions
shown in figure~\ref{fig:Jacobian_poles_1} have chiralities assigned to all pairs of vertices, the splitting of
Riemann spheres terminates at this stage.

Let us finally remark that by giving generic small masses to the internal lines
of the double-box integral with four lightlike external momenta,
we expect that the six spheres that arose in the massless case are turned into a smooth elliptic curve.
It would be interesting to study the situation for general patterns of internal masses.

\subsubsection{Maximal cuts versus integrated expressions}\label{sec:elliptic}

This subsection lies outside the main scope of this paper and could be omitted on a first reading.

Widely propagated folklore holds
that there should be a close connection between the maximal cut of a given integral and the analytic form of its integrated expression.
The appearance of an elliptic curve in case 4 provides closer evidence of such a connection, as we will now argue.
As shown in ref.~\cite{Paulos:2012nu}, eq.~(8.1), the integral in figure~\ref{fig:elliptic10} can be represented as a one-scale integral as follows
\begin{equation}
 I_{(2,2,1,2,2,1)} = \int_u^\infty \frac{du'}{\sqrt{\tilde Q(u')}} \times \left( \textrm{Li}_3(\cdots) +\cdots \right)  \label{mellineq}
\end{equation}
where $\tilde Q(u')$ defines the same elliptic curve as $Q(z)$ in eq.~(\ref{ellipticint}).
It may be shown that the sunrise integral with massive propagators admits a very similar integral representation,
with the integrand of eq.~(\ref{mellineq}) containing $\log(\cdots)$ instead of $\Li_3(\cdots)$ \cite{Caron-Huot:2112}.
This integral was studied analytically in great detail in refs.~\cite{Laporta:2004rb,MullerStach:2011ru} and references therein,
and was found not to be expressible in terms of polylogarithms.
Given the similarity with eq.~(\ref{mellineq}), we thus find it extremely unlikely that $I_{(2,2,1,2,2,1)}$ is expressible in terms of (multiple) polylogarithms.
Thus the topology of the Riemann surface parametrizing the heptacut solutions appears to be reflected in the integrated expression.

Moreover, as we argue in appendix \ref{app:N4}, such more general functions must necessarily be present in the scattering amplitudes of $\mathcal{N}=4$ super Yang-Mills theory.  In this appendix, we point out that a particular two-loop N${}^3$MHV amplitude for the scattering of 10 massless scalars
is given precisely by the integral $I_{(2,2,1,2,2,1)}$ alone.  This suggests that the realm of $\mathcal{N}=4$ SYM extends beyond that of polylogarithms.

\section{Twistor geometry of two-loop maximal cuts}\label{sec:twistors}

Momentum twistor space provides an appealing geometric
rephrasing of the problem of setting massless propagators on-shell.
Accordingly, we devote this section to explain the momentum twistor geometry
of two-loop maximal cuts, the one-loop case having already been
presented in ref.~\cite{Hodges:2010kq}. Besides geometric elegance,
this formulation has the conceptual advantage of being inherently
coordinate free. However, as the results of this section are not used elsewhere in this
paper, this section can be skipped in a first reading.

The main notions are briefly reviewed in section~\ref{sec:twistor_generalities},
after which we describe the twistor space geometry of the heptacut in section~\ref{sec:twistor_heptacut}.

\subsection{Generalities}\label{sec:twistor_generalities}

Given a sequence of null momenta $k_1 , \ldots, k_n$, where
$k_i^\mu = \frac{1}{2} \langle k_i|\gamma^\mu|k_i]$, momentum twistors are
four-component objects defined as \cite{Hodges:2009hk}
\begin{equation}
Z_i^a \equiv \left( \begin{array}{c} \langle k_i| \\ \langle k_i|
\gamma_\mu x_i^\mu \end{array}\right) \hspace{7mm} \mbox{where} \hspace{7mm} x_i
\equiv \sum_{j=1}^{i{-}1} k_j \: . \label{eq:def_of_mom_twistor}
\end{equation}
The definition is invertible: given a configuration of momentum
twistors, a simple and explicitly known formula recovers the
four-momenta $k_i$ \cite{Hodges:2009hk}. An often useful fact is that any
sequence of momentum twistors gives on-shell momenta which respect
momentum conservation. In other words, the momentum twistors are
free variables which solve all phase-space constraints. However, in the
following we will not need this fact, as we will take the momenta
$k_i$ to be given and construct momentum twistors out of them.

In analogy with the spinors $\langle k_i|$, the
momentum twistors $Z_i^a$ are defined only up to an overall
rescaling; that is, they are points in three-dimensional complex
projective space $\mathbb{CP}^3$. Due to this projective invariance,
the subspace spanned by two momentum twistors
$Z_{i-1}$ and $Z_i$ defines a line in momentum twistor space which
we will denote $(i{-}1 \hspace{1mm} i)$.
This two-dimensional span is naturally recorded by the $4\times 2$ matrix $(Z_{i-1} \hspace{0.6mm} Z_i)$.
But since only the two-dimensional subspace itself matters, rather than any basis chosen for it, we can
multiply this matrix from the right by an appropriate GL(2) matrix to obtain (assuming $Z_{i-1} \neq Z_i$) the form
\begin{equation}
 (i{-}1\hspace{1mm}i) \hspace{1mm}\simeq\hspace{1mm} \left(\begin{array}{cc} 1&0 \\ 0&1\\  \multicolumn{2}{c}{\sigma_\mu^{\alpha\dot\beta}x_i^\mu} \end{array}\right) \label{twistorline}
\end{equation}
where $\sigma_\mu^{\alpha\dot\beta}$ are the $2\times 2$ Pauli matrices.
Observe that the line in eq.~(\ref{twistorline})
encodes the value of the associated region momentum $x_i$,
thereby identifying lines in momentum twistor space with points
in region momentum space. Conversely,
given a region momentum, one can construct the matrix in eq.~(\ref{twistorline})
and interpret it as a line in momentum twistor space.

The kinematic invariants that can arise when considering planar graphs take
the form $(k_i+k_{i{+}1}+\cdots+k_j)^2=(x_i - x_{j{+}1})^2$.  An
important fact \cite{Hodges:2009hk}, whose verification we omit here, is that
\begin{equation}
 (k_i+k_{i{+}1}+\cdots+k_j)^2 \langle i{-}1 \hspace{1mm}
 i \rangle\langle j \hspace{1mm} j{+}1\rangle = \langle i{-}1
 \hspace{1mm} i \hspace{1mm} j \hspace{1mm} j{+}1\rangle
 \label{momtwistorsquare}
\end{equation}
where
\begin{equation}
\langle i \hspace{0.5mm} j \hspace{0.5mm} k \hspace{0.5mm}
l\rangle \equiv \epsilon_{abcd}Z_i^aZ_j^bZ_k^cZ_l^d,
\end{equation}
and $\epsilon_{abcd}$ is the antisymmetric Levi-Civita symbol.

What will be important in the following is the geometrical
interpretation of eq.~(\ref{momtwistorsquare}):
geometrically, what eq.~(\ref{momtwistorsquare})
implies is that $(k_i+k_{i{+}1}+\cdots+k_j)^2$ equals zero if
and only if the lines $(i{-}1 \hspace{1mm} i)$ and $(j
\hspace{1mm} j{+}1)$ intersect in momentum twistor space $\mathbb{CP}^3$.
Indeed, if $\langle i{-}1
 \hspace{1mm} i \hspace{1mm} j \hspace{1mm} j{+}1\rangle$ equals zero,
one of the four twistors $Z_{i-1}, Z_i, Z_j, Z_{j+1}$ can be written as a linear combination of the others.
This means that there is some point in $\mathbb{CP}^3$ which is simultaneously
a linear combination of $Z_{i-1}$ and $Z_i$ on one hand, and of $Z_j$ and $Z_{j+1}$ on the other.
That is, the lines $(i-1 \hspace{0.6mm} i)$ and $(j \hspace{0.6mm} j+1)$
intersect.
As noted above, each loop momentum gives rise to a line in $\mathbb{CP}^3$
through its associated region momenta. Thus, when such lines intersect, propagators become on-shell.

\begin{figure}[!h]
\begin{center}
\includegraphics[angle=0, width=0.9\textwidth]{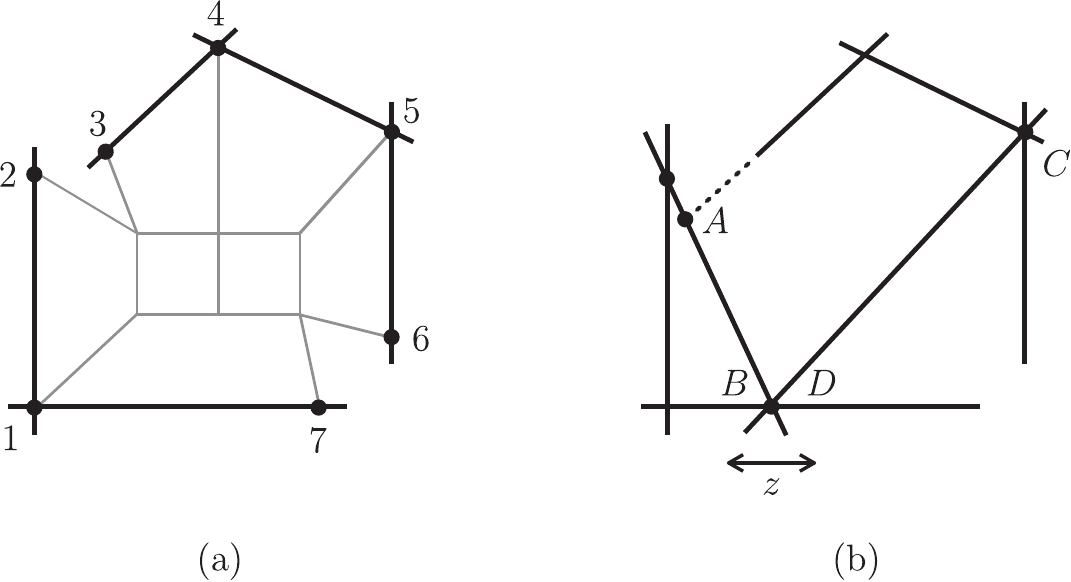}
\caption{(a) The momentum twistor geometry associated with the double-box
topology shown in figure~\ref{fig:Case_1_generic}. Each region
of the exterior of the double-box graph, shown here in gray,
gives rise to a line. The labels refer to the external momenta. (b) The two loop
momenta (lines $(AB)$ and $(CD)$) in the kinematical solution
described in the main text.  The lines $(AB)$ and $(CD)$ are
constrained to intersect each other and the lines in the
background. This solution has a mobile point along the line (71)
corresponding to the free parameter $z$. Note that the points
$(A)$ and $(C)$ are fixed; moreover, $(A)$
is located at the intersection of the plane $(712)$ and the line
$(34)$.} \label{fig:twistor}
\end{center}
\end{figure}

\subsection{Heptacut example}\label{sec:twistor_heptacut}

Let us see how this works for the 7-point topology shown in
figure~\ref{fig:Case_1_generic}.  For conciseness, we will only describe the kinematical solution in which the two three-point vertices in the lower row of the double-box graph
have chirality $\ominus$ and $\oplus$, respectively, and the chirality in the upper-right corner is $\oplus$
(corresponding to the kinematical solution $\mathcal{S}_5$, see figure~\ref{fig:Jacobian_poles_1}).  The other distributions of chiralities are treated in an entirely analogous way.

The first step is to draw the configuration of momentum twistors
associated with the external data, as shown in
figure~\ref{fig:twistor}(a).  The rule is the following: each
region in the exterior of the double-box graph gives rise to a line. Whenever
two regions are separated by a massless external leg of the graph, the
corresponding lines intersect.

The second step is to construct two lines $(AB)$ and $(CD)$, respectively
corresponding to the left and right loop momentum, which have appropriate
intersections with the lines encoding the external momenta. More precisely,
imposing the heptacut constraints, the line $(AB)$ must intersect
with the three lines (71), (12) and (34); and the line $(CD)$ must intersect with (45), (56) and (71).
Furthermore, the lines $(AB)$ and $(CD)$ must intersect each other,
corresponding to having the central propagator cut.
Thus, the problem of setting propagators on-shell translates into a
geometry problem involving straight lines in a three-dimensional space.

Let us solve it in steps. First, $(CD)$ has to intersect $(45)$
and $(56)$. There is one obvious solution: $(CD)$ can pass through
the point $(5)$.  Actually, there is another solution: $(CD)$
could be any line inside the plane $(456)$.  These two discrete
solutions correspond to the $\oplus$ and $\ominus$ vertices,
respectively.  For example, the $\oplus$ vertex joins momenta
whose holomorphic spinors $\langle k_i |$ are proportional,
implying equality of their corresponding momentum twistors
(\ref{eq:def_of_mom_twistor}). Thus, as the upper-right vertex of the double
box is assumed to be $\oplus$, $(CD)$ is a line
passing through the point $(5)$.  Furthermore, $(CD)$ must intersect $(71)$.
Since we have just described two points on the line $(CD)$, and
since only the line they span is meaningful, it is straightforward
to write down a parametrization for $(CD)$,
\begin{equation}
  Z_C = Z_5, \quad Z_D = Z_1+z Z_7
\end{equation}
where $z$ is a free variable.

Next, we consider $(AB)$.  As the lower-left vertex of the double
box is assumed to be $\ominus$, $(AB)$ must be inside the plane
$(712)$. This characterization of $\ominus$ vertices follows from the above characterization of $\oplus$ vertices,
given that parity interchanges points and planes in momentum twistor space \cite{Hodges:2009hk}.
Furthermore, $(AB)$ must intersect with the line $(34)$. This
implies that $(AB)$ must pass through the point where the line $(34)$ intersects
the plane $(712)$. This information allows us to write down the following parametrization
of $(AB)$,
\begin{equation}
  Z_A = \l 7123\r Z_4 - \l 7124\r Z_3, \quad Z_B = Z_1+w Z_7 \: . \label{eq:AB_parametrization}
\end{equation}
Indeed, the coefficients in eq.~(\ref{eq:AB_parametrization}) are such that $\l 712A\r = \l 712B\r = 0$;
that is, the line $(AB)$ is contained in the plane $(712)$.

Finally, what remains to be achieved is to have the lines $(AB)$ and $(CD)$
intersect each other; this is realized by setting $w=z$.  We have thus fully
characterized the solution $\mathcal{S}_5$ as a function of one free
parameter $z$. Through eq.~(\ref{twistorline}), one can proceed to extract
the loop momenta $\ell_1$ and $\ell_2$ as functions of this
variable.

Let us now consider the behavior of the solution as a
function of $z$ (see figure~\ref{fig:twistor}(b)). For instance, as
$z\to 0$, $(B)$ approaches the point $(1)$, and the lower-left
vertex of the double-box graph becomes simultaneously $\oplus$ and $\ominus$. Thus
$\mathcal{S}_5$ merges with $\mathcal{S}_2$.  Similarly, there is
a point ($z=-\frac{\l4561\r}{\l4567\r}$) where the line $(CD)$ becomes
supported inside the plane $(456)$.  At that point, the
upper-right vertex becomes simultaneously $\oplus$ and $\ominus$,
and $\mathcal{S}_5$ merges with $\mathcal{S}_3$.

Using only such geometrical reasoning, one may construct all the kinematical solutions
to the heptacut constraints of a given double-box integral
and proceed to prove that they link into a chain,
as already discussed and proven in sections~\ref{sec:Case_1}-\ref{sec:Case_4}.

\section{Uniqueness of two-loop master contours}\label{sec:proliferation_of_contours}

In reference \cite{Kosower:2011ty}, the first steps were taken in extending
maximal unitarity beyond one loop. It was argued there that the contours
of integration associated with two-loop maximal cuts cannot be chosen
arbitrarily; rather, they are subject to the consistency condition that
any function which integrates to zero on the real slice
$\mathbb{R}^D \times \mathbb{R}^D$ (where $D=4-2\epsilon$)
must also integrate to zero on the maximal-cut contour.
This constraint on the contour ensures that two Feynman integrals
which are equal, possibly through some non-trivial relations, will
also have identical maximal cuts. As argued in ref. \cite{Kosower:2011ty},
contours satisfying this consistency condition are guaranteed to
produce correct results for scattering amplitudes in any quantum field theory.
To be more accurate, the validity of the approach extends to all quantum field theories in which all states are massless.

In particular, no assumptions need to be made regarding the powers of loop momentum present in numerators. This is due to the nontrivial, but empirically true fact that the number of master integrals in a given topology is independent of the loop-momentum power counting beyond a certain threshold (see, for instance, ref.~\cite{Gluza:2010ws}).
Therefore, this method can be applied indiscriminately to $\mathcal{N}=4,2,1,0$ Yang-Mills theory, massless QCD, gravity or scalar theories (in the latter two cases, as soon as the analysis has been extended beyond the planar sector).

The maximal cut of the double-box integral is in general a contour integral
in the complex plane.
For the case of four massless external momenta $k_1, \ldots, k_4$,
it turns out that the contour may encircle 14 different leading singularities
with some a priori undetermined winding numbers, in the notation
of ref.~\cite{Kosower:2011ty} denoted as\footnote{Please note that ref.
\cite{Kosower:2011ty} employed a rescaled version of the loop momentum parametrization
(\ref{eq:l1_parametrization})-(\ref{eq:l2_parametrization}) used here.
As a result of the rescaling, all nonzero Jacobian poles are located at
$z=-\chi$.}
\begin{eqnarray}
a_{1,i} &\longrightarrow& \mathrm{encircling} \hspace{2mm} z=0       \hspace{2mm} \mbox{for solution} \hspace{1.2mm} \mathcal{S}_i \nonumber \\
a_{2,i} &\longrightarrow& \mathrm{encircling} \hspace{2mm} z=-\chi   \hspace{2mm} \mbox{for solution} \hspace{1.2mm} \mathcal{S}_i \label{eq:notation_for_winding numbers}\\
a_{3,j} &\longrightarrow& \mathrm{encircling} \hspace{2mm} z=-\chi-1 \hspace{2mm} \mbox{for solution} \hspace{1.2mm} \mathcal{S}_j \nonumber
\end{eqnarray}
where $i=1,\ldots,6$, $j=5,6$ and $\chi \equiv \frac{s_{14}}{s_{12}}$.
Here, the 12 winding numbers $a_{1,i}$ and $a_{2,i}$ are associated with Jacobian poles
whereas the 2 winding numbers $a_{3,5}$ and $a_{3,6}$ are associated with the
poles $\infty_R$ in $\mathcal{S}_5$ and $\mathcal{S}_6$ in figure~\ref{fig:Jacobian_poles_1}.
The consistency conditions on the maximal-cut contours were shown
to translate into the following linear constraints on the
winding numbers,
\begin{equation}
\begin{array}{rll}
a_{1,2} + a_{1,5} - a_{1,4} - a_{1,6}  &=&  0  \\[0.5mm]
a_{2,1} + a_{2,2} - a_{2,3} - a_{2,4}  &=&  0  \\[0.5mm]
a_{2,6} - a_{1,1} - a_{2,5} + a_{1,3}  &=&  0  \\[0.5mm]
a_{3,5} - a_{3,6}                      &=&  0  \\[0.5mm]
a_{1,2} + a_{1,5} + a_{1,4} + a_{1,6}  &=&  -a_{2,6} + a_{1,1} - a_{2,5} + a_{1,3} + a_{3,5} + a_{3,6}  \\[0.5mm]
a_{3,5} + a_{3,6}                      &=&  -\frac{1}{2} \sum_{j=1}^6 (a_{1,j} - a_{2,j}) + \frac{3}{2} \sum_{j\neq1,3} a_{1,j}
\end{array}  \label{eq:contour_eqs}
\end{equation}
where the first four follow from the vanishing integrations of
integrands involving Levi-Civita contractions of loop momenta
and the last two follow from integration-by-parts identities
between tensor double-box integrals. Imposing these constraints
on the winding numbers leaves $14 - 6 = 8$ free parameters
in the contours.

On the other hand, in the case of four massless external momenta,
there turn out to be exactly two linearly independent master integrals
of the double-box topology. The heptacuts of the particular
masters used in ref.~\cite{Kosower:2011ty} were found there to evaluate to
\begin{eqnarray}
\chi s_{12}^3 I^{\hspace{0.5mm} \mathrm{cut}}_{1,1,0,1,1,0} [1]  &=& \sum_{i=1}^6 (a_{1,i} - a_{2,i})        \label{eq:heptacut_scalar_DB_old} \\
2 s_{12}^2 I^{\hspace{0.5mm} \mathrm{cut}}_{1,1,0,1,1,0} [\ell_1 \cdot k_4]  &=&  \sum_{i \neq 1,3} a_{1,i}  \label{eq:heptacut_tensor_DB_old}
\end{eqnarray}
where we remind the reader of the numerator insertion notation explained in eq.~(\ref{eq:def_double_box}).
As shown in detail in this reference, it is possible to find
contours $(a_{1,i}, a_{2,i}, a_{3,j})$ satisfying the constraint
equations (\ref{eq:contour_eqs}) with the additional property of setting
the right hand side of eq.~(\ref{eq:heptacut_tensor_DB_old}) to zero
while setting the right hand side of eq.~(\ref{eq:heptacut_scalar_DB_old}) to one (or vice versa).
Such contours thus isolate the contribution of a single
master integral in the basis decomposition (\ref{eq:2-loop_basis_decomposition}) and are
therefore referred to as \emph{master contours}. With the 8 free parameters in the
contours that remained in the previous paragraph, we thus find a total of $8 - 2 = 6$
free parameters in the two-loop master contours.

Extrapolating from the situation at one loop in which there is a unique
contour associated with each of the basis integrals \cite{Britto:2004nc,Forde:2007mi},
we would not expect any free parameters in the master contours, and the appearance of 6 unconstrained variables comes as a surprise.
\\
\\
By considering figure~\ref{fig:Jacobian_poles_1}, this phenomenon can now
easily be explained: all of the Jacobian poles belong to
two Riemann spheres and so are counted twice in the above
counting (\ref{eq:notation_for_winding numbers}).
Indeed, for example, the winding numbers $a_{1,2}$ and $a_{1,5}$
only arise in the combination $(a_{1,2}+a_{1,5})$, as is visible in eq.~(\ref{eq:contour_eqs}).
Due to this, a contour which encircles the $z=0$ pole in $\mathcal{S}_2$ and in $\mathcal{S}_5$
with identical winding numbers in the two spheres, and no other poles, is equivalent to
a zero-cycle. The addition of such zero-cycles defines an equivalence
relation on the vector space spanned by the leading-singularity contours: any two
contours related by the addition of such zero-cycles are equivalent.

\clearpage

This manifests itself as the invariance of the contour constraint equations (\ref{eq:contour_eqs})
under the translations\footnote{Please note that in this section,
in order to make the connection with ref.~\cite{Kosower:2011ty} as clear as possible,
we adopt the conventions of this reference on the orientations of contours encircling
poles in the various Riemann spheres and on the signs of the Jacobians. These conventions
differ from those used in this paper, in particular, with respect to the minus signs described below eq.~(\ref{hexacut}).
The pattern of relative signs between the zero-cycle winding numbers $(\xi_i, \pm \xi_i)$
in eqs. (\ref{eq:contour_eq_invariance_1})-(\ref{eq:contour_eq_invariance_6}) owes to the omission of these minus signs
in ref.~\cite{Kosower:2011ty}.}
\begin{align}
(a_{2,1}, \hspace{0.4mm} a_{2,2})  \hspace{1.8mm} &\longrightarrow \hspace{1.8mm}  (a_{2,1}, \hspace{0.4mm} a_{2,2}) \hspace{0.4mm}+\hspace{0.4mm} (\xi_1, -\xi_1)  \label{eq:contour_eq_invariance_1}\\
(a_{1,2}, \hspace{0.4mm} a_{1,5})  \hspace{1.8mm} &\longrightarrow \hspace{1.8mm}  (a_{1,2}, \hspace{0.4mm} a_{1,5}) \hspace{0.4mm}+\hspace{0.4mm} (\xi_2, -\xi_2)  \label{eq:contour_eq_invariance_2}\\
(a_{2,5}, \hspace{0.4mm} a_{1,3})  \hspace{1.8mm} &\longrightarrow \hspace{1.8mm}  (a_{2,5}, \hspace{0.4mm} a_{1,3}) \hspace{0.4mm}+\hspace{0.4mm} (\xi_3, \xi_3)   \label{eq:contour_eq_invariance_3}\\
(a_{2,3}, \hspace{0.4mm} a_{2,4})  \hspace{1.8mm} &\longrightarrow \hspace{1.8mm}  (a_{2,3}, \hspace{0.4mm} a_{2,4}) \hspace{0.4mm}+\hspace{0.4mm} (\xi_4, -\xi_4)  \label{eq:contour_eq_invariance_4}\\
(a_{1,4}, \hspace{0.4mm} a_{1,6})  \hspace{1.8mm} &\longrightarrow \hspace{1.8mm}  (a_{1,4}, \hspace{0.4mm} a_{1,6}) \hspace{0.4mm}+\hspace{0.4mm} (\xi_5, -\xi_5)  \label{eq:contour_eq_invariance_5}\\
(a_{2,6}, \hspace{0.4mm} a_{1,1})  \hspace{1.8mm} &\longrightarrow \hspace{1.8mm}  (a_{2,6}, \hspace{0.4mm} a_{1,1}) \hspace{0.4mm}+\hspace{0.4mm} (\xi_6, \xi_6) \:,
\label{eq:contour_eq_invariance_6}
\end{align}
corresponding to the addition of a zero-cycle encircling
each Jacobian pole with the winding numbers $(\xi_i, \pm \xi_i) \in \mathbb{Z} \times \mathbb{Z}$
on the two Riemann spheres containing the pole. These equivalence relations
allow us to add to an arbitrary contour, characterized
by the winding numbers $(a_{1,i}, a_{2,i}, a_{3,j})$, the
zero-cycle with $(\xi_i) = (-a_{2,1}, -a_{1,2}, -a_{2,5}, -a_{2,3}, -a_{1,4}, -a_{2,6})$
to obtain an equivalent contour characterized by 8 independent parameters.
This shows that the leading singularity contours are characterized
by 8 rather than 14 winding numbers.

\subsection{Invariant labeling of contours}

To get around the redundancy built into the
notation~(\ref{eq:notation_for_winding numbers}),
we will from now on adopt the following notation for the independent winding numbers
\begin{equation}
\Omega = (\omega_{1\cap2}, \hspace{0.7mm} \omega_{2\cap5}, \hspace{0.7mm} \omega_{5\cap3}, \hspace{0.7mm}
\omega_{3\cap 4}, \hspace{0.7mm} \omega_{4\cap6}, \hspace{0.7mm} \omega_{6\cap 1}, \hspace{0.7mm}
\omega_{5,\infty_R}, \hspace{0.7mm} \omega_{6,\infty_R}) \: . \label{eq:new_notation_for_winding_numbers}
\end{equation}
Here $\omega_{i \cap j}$ denotes the winding around the
intersection point of $\mathcal{S}_i$ and $\mathcal{S}_j$
of a small circle supported in either of these spheres, with positive orientation
in $\mathcal{S}_i$ or with negative orientation in $\mathcal{S}_j$. Analogously,
$\omega_{j,\infty_R}$ denotes the winding around the point $\infty_R$
in $\mathcal{S}_j$ where the right loop momentum $\ell_2$ becomes infinite (see figure~\ref{fig:Jacobian_poles_1}).
Of course, we could trade some of the variables in
eq.~(\ref{eq:new_notation_for_winding_numbers}) for winding
numbers around the other infinity poles in figure~\ref{fig:Jacobian_poles_1},
but we find the above choice to be the most convenient.
Also note that (\ref{eq:new_notation_for_winding_numbers})
has the added advantage over the notation~(\ref{eq:notation_for_winding numbers})
of not making reference to a particular parametrization
of the Riemann spheres $\mathcal{S}_i$, hence the title of this subsection.

The 8 winding numbers $\omega_i$ in eq.~(\ref{eq:new_notation_for_winding_numbers}) are equal
to the following linear combinations of the $a_{i,j}$
\begin{equation}
\Omega = (a_{2,1}+a_{2,2}, \hspace{0.7mm} -a_{1,2}-a_{1,5}, \hspace{0.7mm} a_{2,5}-a_{1,3}, \hspace{0.7mm}
a_{2,3}+a_{2,4}, \hspace{0.7mm} -a_{1,4}-a_{1,6}, \hspace{0.7mm} a_{2,6}-a_{1,1}, \hspace{0.7mm} a_{3,5}, \hspace{0.7mm} a_{3,6}) \: .
\label{eq:def_of_Omega}
\end{equation}
In analogy with the notation~(\ref{eq:new_notation_for_winding_numbers})
for the winding numbers around the leading singularities, let us introduce
the following notation for the residues. The residue at the intersection
point of the spheres $\mathcal{S}_i$ and $\mathcal{S}_j$,
computed from the viewpoint of sphere $\SS_i$, will be
labeled $\Res_{i\cap j}$. Alternatively, we could consider the same residue,
but computed from the viewpoint of $\SS_j$.
As noted around eqs.~(\ref{eq:residues_of_cuts_relation_1})-(\ref{eq:residues_of_cuts_relation_6}), the result would be equal
and opposite. Thus, we can re-express these identities in a very compact form
by declaring $\Res_{i \cap j}$ to be antisymmetric,
\be
  \Res_{i\cap j} \Phi (z) = -\Res_{j\cap i} \Phi (z) \: . \label{eq:residues_of_cuts_Res}
\ee
Here $\Phi$ denotes an arbitrary function of the loop momenta.
The other residues will be labeled $\Res_{i,\infty_L}$ and $\Res_{i,
\infty_R}$, according to either the left or right loop momentum of the
double-box graph approaching infinity. The identities~(\ref{eq:residues_of_cuts_relation_1})-(\ref{eq:residues_of_cuts_relation_6})
are dual to the contour equivalence relations~(\ref{eq:contour_eq_invariance_1})-(\ref{eq:contour_eq_invariance_6});
an application of them is given in appendix~\ref{sec:cuts_across_kin_sols}.
\\
\\
Re-expressing the contour constraint equations~(\ref{eq:contour_eqs}) in terms of the
winding numbers (\ref{eq:new_notation_for_winding_numbers})-(\ref{eq:def_of_Omega}), they are found to take the form
\begin{equation}
\begin{array}{rll}
 \omega_{2 \cap 5}   -  \omega_{4 \cap 6}    &=&  0  \\[0.5mm]
 \omega_{1 \cap 2}   -  \omega_{3 \cap 4}    &=&  0  \\[0.5mm]
\omega_{6 \cap 1}   -  \omega_{5 \cap 3}    &=&  0  \\[0.5mm]
\omega_{5,\infty_R}  -  \omega_{6,\infty_R}  &=&  0  \\[0.5mm]
\omega_{2 \cap 5}   +  \omega_{4 \cap 6}   &=&  \omega_{5 \cap 3} + \omega_{6 \cap 1}  - \omega_{5,\infty_R} - \omega_{6,\infty_R}  \\[0.5mm]
\omega_{5,\infty_R}  +  \omega_{6,\infty_R}  &=&  \frac12(\omega_{1 \cap 2} + \omega_{3 \cap 4}+\omega_{5 \cap 3}+ \omega_{6 \cap 1}) -\omega_{2 \cap 5} - \omega_{4 \cap 6}
\end{array}  \label{eq:relabeled_contour_eqs}
\end{equation}
while the heptacut master double boxes used in ref.~\cite{Kosower:2011ty}
evaluate to
\begin{eqnarray}
\chi s_{12}^3 I^{\hspace{0.5mm} \mathrm{cut}}_{1,1,0,1,1,0} [1]    &=&  \omega_{1 \cap 2} +\omega_{2 \cap 5} + \omega_{5 \cap 3} +  \omega_{3 \cap 4}+ \omega_{4 \cap 6} + \omega_{6 \cap 1} \label{eq:heptacut_scalar_DB} \\
2 s_{12}^2 I^{\hspace{0.5mm} \mathrm{cut}}_{1,1,0,1,1,0} [\ell_1 \cdot k_4]  &=&  \omega_{2 \cap 5} + \omega_{4 \cap 6} \, . \label{eq:heptacut_tensor_DB}
\end{eqnarray}
We observe that upon imposing the 6 constraint equations~(\ref{eq:relabeled_contour_eqs})
on the 8 winding numbers $\omega_{i\cap j}$ given in eq.~(\ref{eq:new_notation_for_winding_numbers}),
we are left with 2 unconstrained parameters. This number of free parameters exactly
equals the number of master double-box integrals at four points.
In other words, we observe that in terms of the $\omega$-variables,
there is a unique master contour associated with each of the master
double-box integrals in eqs.~(\ref{eq:heptacut_scalar_DB})-(\ref{eq:heptacut_tensor_DB}).
These contours are respectively characterized by the winding numbers
\begin{equation}
\Omega_1=\frac14(1,0,1,1,0,1,1,1) \hspace{7mm} \mbox{and} \hspace{7mm} \Omega_2=-\frac14(1,-2,1,1,-2,1,3,3) \: .
\label{eq:scalar_tensor_master_contours}
\end{equation}
The observed uniqueness of master contours at two loops
is in perfect analogy with the situation in one-loop generalized unitarity \cite{Britto:2004nc,Forde:2007mi}
and constitutes the main result of this section.

\subsection{A basis with infrared finite master integrals?} \label{sec:irfinite}

As indicated in the basis decomposition of two-loop amplitudes in eq.~(\ref{eq:2-loop_basis_decomposition}),
the integral coefficients are functions of the dimensional regulator $\epsilon$.
But in contrast to the situation at one loop, the
$\mathcal{O}(\epsilon)$ contributions to the coefficients
cannot be re-expressed as rational contributions to the
amplitude, and these corrections therefore form
an inevitable part of the two-loop integral coefficients.
The physical significance of these $\mathcal{O}(\epsilon)$
contributions lies in the fact that in the basis expansion
of the two-loop amplitude, they will multiply
$\frac{1}{\epsilon^k}$ singularities in the integrated
expressions for the two-loop integrals, thus producing
finite contributions to the amplitude. Their extraction
therefore poses an important problem.
Unfortunately, as explained in ref.~\cite{Bern:2000dn},
the $\mathcal{O}(\epsilon)$ parts of two-loop integral coefficients
are not obtainable from four-dimensional cuts. Instead,
they must be computed by evaluating cuts in $D=4-2\epsilon$ dimensions,
something which is technically much more involved.

Ideally, one would like to circumvent the need for taking cuts
in $D=4-2\epsilon$ dimensions, or at least limit such computations
as much as possible. One way to achieve this would be to expand the two-loop amplitude
in a basis that contains as many infrared finite integrals as possible.
Indeed, although the expansion coefficients may still depend on $\epsilon$,
the physically relevant part of the coefficients multiplying IR finite integrals
is purely $\mathcal{O}(\epsilon^0)$ and may thus be obtained from
strictly four-dimensional cuts. Of course, as two-loop amplitudes
do have IR divergences of their own (as they necessarily must, to cancel
the IR divergences of tree and one-loop amplitudes in the cross-section \cite{Kinoshita:1962ur,Lee:1964is}),
any basis of integrals must contain IR divergent integrals\footnote{Unless
we are prepared to accept integral coefficients with $\frac{1}{\epsilon^k}$ singularities
-- which we are not.}. Nonetheless, it is plausible that by using
a basis with a minimal number of IR divergent integrals
one can minimize the work of extracting the physically relevant part
of the basis integral coefficients. Focusing our attention to the double-box contributions to two-loop
amplitudes, we thus turn to the question: can one find two linearly independent
infrared finite integrals with the double-box topology?
\\
\\
A class of integrals with the property of infrared finiteness was introduced
in ref.~\cite{ArkaniHamed:2010gh} where they were used to express the
integrand of $\mathcal{N}=4$ super Yang-Mills amplitudes in a strikingly simple form.
Here, we wish to investigate whether these so-called chiral numerator integrals
can be used as master integrals for two-loop amplitudes in \emph{any} gauge theory.

For four lightlike external momenta there are, up to parity
conjugation, two distinct chiral numerator integrals, which we can define as\footnote{We
should stress that the chiral numerator integrals considered here are not exactly chiral in the sense of ref.~\cite{ArkaniHamed:2010gh},
in which the general notion of a chiral integral was formulated in terms of the analytic structure
of its integrand.  There, chiral integrals were defined
as integrals which have at most simple-pole singularities on each leading-singularity
contour, and whose residues are all equal to either zero, or plus or minus one.
In general, such integrals may differ from the chiral numerator integrals $I_{+ \pm}$
considered here by the addition of integrals with fewer internal lines, such as triangle-boxes or double-triangles.  Since
we are not concerned with the latter integrals in this paper, this distinction will not
be of relevance here, however.}
\ba
 I_{++} &\equiv& I_{1,1,0,1,1,0} \big[ [1 | \slashed{\ell}_1 | 2 \rangle \langle 3 | \slashed{\ell}_2 | 4 ] \big]   \hspace{3mm}\times\hspace{3mm}  [2 \hspace{0.7mm} 3]\langle 1 \hspace{0.7mm} 4\rangle \label{def_chiral_integrals_1} \\
 I_{+-}  &\equiv& I_{1,1,0,1,1,0} \big[ [1 | \slashed{\ell}_1 | 2 \rangle \langle 4 | \slashed{\ell}_2 | 3 ] \big]   \hspace{3mm}\times\hspace{3mm}  [2 \hspace{0.7mm} 4]\langle 1 \hspace{0.7mm} 3\rangle  \label{def_chiral_integrals_2}
\ea
where we remind the reader of the numerator insertion notation explained in eq.~(\ref{eq:def_double_box}).
Similar definitions can be given for an arbitrary number of external legs, replacing the spinors $[i|$ or $\langle i|$ by their flattened
counterparts $[K_i^\flat|$ or $\langle K_i^\flat|$.

The maximal cuts of these integrals are found to take the form
\begin{align}
{I_{++}}^{\hspace{-2.7mm} \mathrm{cut}} &=\hspace{0.8mm} \omega_{6,\infty_R}
 \label{eq:4_point_chiral_int_1} \\
{I_{+-}}^{\hspace{-2.7mm} \mathrm{cut}} &=\hspace{0.8mm} \omega_{1\cap 2} \: ,
\end{align}
receiving contributions from a remarkably small number of leading singularities.
The vanishing of a large number of leading-singularity residues reflects the absence of infrared singularities in the uncut integrals,
as observed in section~\ref{sec:Case_1}. Here, we could of course also have chosen to consider the
integrals $I_{--}$ and $I_{-+}$ whose numerator insertions are obtained by parity
conjugation $\langle \cdot | \cdot | \cdot ] \longleftrightarrow
[\cdot | \cdot | \cdot \rangle$ of
eqs.~(\ref{def_chiral_integrals_1})-(\ref{def_chiral_integrals_2}).
However, on the solution of the first four constraint equations
of~(\ref{eq:relabeled_contour_eqs}) (which express parity invariance of the
contours), one finds that the maximal cuts of parity-conjugate
integrals are equal. The largest potentially linearly independent set of
chiral double boxes therefore consists,
at four points, of two integrals which we choose as those given
in eqs.~(\ref{def_chiral_integrals_1})-(\ref{def_chiral_integrals_2}).

Considering these two
integrals, let us now ask: are they linearly independent as master integrals?
Equivalently, can one find two distinct contours, satisfying the constraint
equations~(\ref{eq:relabeled_contour_eqs}), each with the property of
yielding a nonvanishing maximal cut for precisely one of these
integrals?

Remarkably, such master contours do exist: the contours which extract the coefficient of $I_{++}$ and $I_{+-}$ are, respectively,
\begin{equation}
\Omega_{++}=(0,-1,0,0,-1,0,1,1) \hspace{7mm} \mbox{and} \hspace{7mm} \Omega_{+-}=(1,1,1,1,1,1,0,0) \: . \label{eq:chiral_master_contours}
\end{equation}
Thus the integrals $I_{++}$ and $I_{+-}$ are
linearly independent and may be used as master integrals
for the double-box contributions to two-loop amplitudes in \emph{any} gauge theory.
Incidentally, as a bonus, the contours in eq.~(\ref{eq:chiral_master_contours}) take a somewhat simpler form than
those in eq.~(\ref{eq:scalar_tensor_master_contours})
whose winding numbers display no easily discernible pattern.
\\
\\
The choice of using the chiral numerator integrals as master integrals
provides a substantial simplification over other choices
(such as that of eqs.~(\ref{eq:heptacut_scalar_DB})-(\ref{eq:heptacut_tensor_DB})),
as their infrared finiteness allows their expansion coefficients in
the two-loop amplitude to be obtained from strictly four-dimensional cuts.
But one might worry that the technical difficulty is simply shifted
elsewhere, in particular to the analytical evaluation of these integrals.
To counter this concern, we present in the next section a detailed analytical
evaluation of these integrals and moreover find the result to take a remarkably compact form.

\section{Analytical evaluation of chiral double boxes}\label{sec:analytical_evaluation_of_chiral_DBs}

In the previous section we found that the chiral numerator integrals
in eqs.~(\ref{def_chiral_integrals_1})-(\ref{def_chiral_integrals_2}) form a basis of the master integrals with the double-box topology.
The finiteness of these integrals allows for an economic extraction of their coefficients, which can be done directly in four space-time dimensions.
In this section we turn to the question of evaluating these integrals analytically.
We will focus our attention on obtaining analytical expressions in the special
case of four lightlike external momenta, but we are hopeful that many
of the ideas presented here will prove applicable for higher
numbers of external legs as well.

Our procedure goes through several steps. First, we apply to the
Feynman parametrized expression of the integrals a sequence of
non-obvious changes of variables, motivated by recent work
\cite{Paulos:2012nu} on Mellin space transforms, but which make no reference to Mellin space.
This enables us to obtain the symbols \cite{Goncharov:2010jf,Duhr:2011zq} of each of the chiral double boxes.
The finiteness of the integrals, ensuring that they can be evaluated directly
in four dimensions, is very helpful in this respect.
Finally, imposing the constraints of uniform transcendentality,
analyticity and Regge limits then allow us to unambiguously
integrate these symbols, leaving us with the final expressions in eqs.~(\ref{eq:chiral_DBs_final_result_1})-(\ref{eq:chiral_DBs_final_result_2}).

Our starting point will be the (standard) Feynman parametrization
formula, which gives for a double box with arbitrary numbers of legs\footnote{We use the
normalization $I[\cdots]\equiv \int \frac{d^D\ell_1 d^D\ell_2
(\cdots)}{(i \pi^{D/2})^2 \mbox{(inverse propagators)}}$ where $D
= 4 + 2\epsilon$ is the spacetime dimension, here set directly equal to 4.}\ba
 I_{i_1, i_2, i_3, i_4, i_5, i_6}[ v_1{\cdot}\ell_1 ~ v_2{\cdot}\ell_2]
&=& \int_0^1 da_1\cdots da_3 \hspace{0.9mm} db_4\cdots db_6
\hspace{0.9mm} dc \hspace{0.9mm} c \hspace{0.9mm}
\delta\Big(1-\sum_i a_i-\sum_i b_i -c \Big) \hspace{0.8mm}
\mathcal{U}^{-1} \nl && \hspace{1.5cm} \times \left(
 \frac{2c\big(v_1{\cdot}\sum_i b_i x_{i3} \big)
 \big(v_2{\cdot}\sum_i a_i x_{i6} \big)}{\mathcal{V}^3}
-\frac{v_1{\cdot}v_2}{\mathcal{V}^2}\right) \nl
 &\equiv& I_2- v_1{\cdot}v_2I_1  \label{Feynman}
\ea
where
\begin{align}
\mathcal{U} &=\sum_i a_i \sum_i b_i + \Big(\sum_i
a_i+\sum_i b_i \Big)c \nonumber \\
 \mathcal{V} &= \Big(\sum_{i<j}a_ia_j x_{ij}^2 \Big) \Big( c+\sum_kb_k \Big)
 + \Big(\sum_{i,j}a_ib_jx_{ij}^2 \Big)c + \Big(\sum_{i<j}b_ib_j x_{ij}^2 \Big)
 \Big(c+\sum_ka_k \Big),
\end{align}
and $x_{ij}\equiv k_i+\cdots+k_{j-1}$. The
present notation is consistent with the dual coordinates $x_i$
introduced in section \ref{sec:twistors}, setting $x_{ij}\equiv
x_i-x_j$. In eq.~(\ref{Feynman}), we have dropped a number of
terms of the form $v_1{\cdot}k_1$ or $v_1{\cdot}k_2$, which vanish
when $v_1, v_2$ are chosen to correspond to the chiral numerators
defined below eqs.~(\ref{def_chiral_integrals_1})-(\ref{def_chiral_integrals_2}).

 In the special case of four lightlike external momenta, the second term
on the last line of eq. (\ref{Feynman}) takes the form \be
 I_1(\chi) = \int \frac{d^3a \hspace{0.8mm} d^3b \hspace{0.8mm}
 dc \hspace{0.8mm} c \hspace{0.8mm} \delta\big(1-c-\sum_i a_i -\sum_i b_i
 \big) \Big(\sum_i a_i \sum_i b_i + c \big(\sum_i a_i +\sum_i b_i \big) \Big)^{-1}}
 {\Big( a_1a_3 \big(c+\sum_i b_i \big) + (a_1b_4+a_3b_6+a_2b_5\chi)c
 + b_4b_6 \big(c+\sum_i a_i \big) \Big)^2}  \label{I1v1}
\ee with $\chi\equiv \frac{t}{s}\equiv \frac{s_{14}}{s_{12}}$.
Furthermore, it turns out that \be I_2(\chi) \hspace{1mm} \propto
\hspace{1mm} \frac{\partial}{\partial \chi} I_1 (\chi) \: . \label{eq:I2_propto_deriv_of_I1} \ee This can
be seen from eq.~(\ref{Feynman}), because $x_{34}=0$ in this case
(that is, $K_3=0$) and $v_1{\cdot} x_{63}=0$, so only the $b_5$
term in the $v_1{\cdot}\sum_i b_i x_{i3}$ factor contributes.  Due to similar simplifications in the $v_2$ factor, the first term
is proportional to $a_2b_5$ which is what the derivative produces.
Plugging in the explicit expressions for $v_1$ and $v_2$ and including the appropriate
constant of proportionality in eq. (\ref{eq:I2_propto_deriv_of_I1}), one finds
\begin{eqnarray}
 I_{++} &=&  -\chi^2 \left( 1+ (1+\chi)\frac{\partial}{\partial\chi}\right) I_1(\chi) \label{I++1} \quad \mbox{and} \nl
 I_{+-} &=& -(1+\chi)^2 \left( 1 + \chi \frac{\partial}{\partial\chi} \right) I_1(\chi) \: .
\end{eqnarray}
The upshot is that we only have to compute $I_1$.
This particular simplification is probably specific to four points, but we welcome it.

\subsection{Evaluation of $I_1$}

It turns out to be possible to evaluate $I_1$ analytically, as we will now describe.

\subsubsection{Step 1: Projectivization}

\def\II{\mathcal{I}}

We would prefer a form for $I_1$ in which such high powers of the integration variables do not occur within the denominator.
This can be obtained using the following trick, which is not limited to the present integral.
This might be called a ``projectivization'' trick.

Thus, our goal in this subsection will be to derive the following equivalent form of eq. (\ref{I1v1}):
\be
 I_1(\chi)= 6\int_1^\infty dc \int_0^\infty \frac{d^7(a_1a_2a_3a_\II b_1b_2b_3b_\II)}{\textrm{vol(GL(1))}} \frac{1}{(cA^2+A.B+B^2)^4} \label{I1v2}
\ee where $A^2\equiv a_1a_3 + a_\II (a_1+a_2+a_3)$, $A.B\equiv
b_1b_3+a_\II (b_1+b_2+b_3)+b_\II (a_1+a_2+a_3)+a_2b_2 \chi$ and
$B^2\equiv b_1b_3+b_\II (b_1+b_2+b_3)$. The ``$1/\textrm{vol(GL(1))}$''
notation is an instruction to set any one of the variables
$a_1, \ldots, b_\II$ equal to 1 and integrate over the seven
remaining ones; this is also why we wrote $``d^7"$ instead of
$``d^8"$. As the notation suggests, the result does not depend on
which variable is set to 1, due to the scaling (``GL(1)'') symmetry of the integrand.

The reader not interested in following the derivation
of eq. (\ref{I1v2}) from eq. (\ref{I1v1}) can safely skip
to the next subsection.

The ``trick'' here is just a sequence of elementary, if
non-obvious, changes of variables. First we use the
$\delta$-function in eq. (\ref{I1v1}) to perform for instance the $a_1$ integration, and
make the change of variables\footnote{This is actually a
composition of two simple changes of variables: first $(a_i,b_i,c)\to
(a_i,b_i,c)/\big(c+\sum_i a_i +\sum_i b_i \big)$ to go from Feynman parameters to projectively identified Schwinger parameters, followed by the rescaling
$b_i\to b_i c/\big(c+\sum_i a_i \big)$.}: \be
 a_i \to \frac{a_i \big(c+\sum_i a_i \big)}{\big(c+\sum_i a_i \big)^2 + c \sum_i b_i} \: ,
 \quad b_i\to \frac{b_i c}{\big(c+\sum_i a_i \big)^2 + c \sum_i b_i} \: ,
 \quad c\to \frac{c \big(c+\sum_i a_i \big)}{\big( c+\sum_i a_i \big)^2 + c \sum_i b_i} \: .
\ee
This brings eq.~(\ref{I1v1}), after a convenient relabeling $(b_4,b_5,b_6)\to (b_3,b_2,b_1)$, into the form
\be
 I_1(\chi) = \int_0^\infty \frac{d^6(a_ib_ic)}{\mbox{vol(GL(1))}}
 \frac{1}{c\big(\sum_i a_i + \sum_i b_i \big) \big( c+\sum_i a_i \big)
 (a_1a_3 c' + a_1 b_3 + a_3 b_1 + a_2 b_2\chi + b_1 b_3)^2}
\ee where $c'=1+ \frac{\sum_i a_i + \sum_i b_i}{c}$. As the
notation suggests, the result does not depend on which variable
was used to solve the $\delta$-function constraint. Second, we
change integration variable $c\to c'$ and then remove the prime, \ba
 I_1(\chi) &=&\int_1^\infty dc \int \frac{d^5(a_ib_i)}{\mbox{vol(GL(1))}}
 \frac{1}{\big(\sum_i a_i + \sum_i b_i \big) \big( c\sum_i a_i
 +\sum_i b_i \big)} \nl &&
 \hspace{24mm} \times \frac{1}{(a_1a_3 c + a_1 b_3 + a_3 b_1 + a_2 b_2\chi + b_1 b_3)^2} \: .
 \label{I1v1c}
\ea
Finally, we combine the three denominators in eq. (\ref{I1v1c}) into one by
introducing the two auxiliary Feynman parameters $a_\II$ and $b_\II$; these are such that integrating out
$a_\II$ and $b_\II$ from eq.~(\ref{I1v2}) produces eq.~(\ref{I1v1c}).
(In a momentum twistor space computation, $a_\II$ and $b_\II$ would have appeared automatically as Feynman parameters for the propagators $\l AB\mathcal{I} \r$ and $\l CD\mathcal{I}\r$, connecting to the so-called infinity point $\mathcal{I}$. This is the origin of our notation $a_\II, b_\II$.)

Our desire to obtain the form in eq.~(\ref{I1v2}) was sparked by a recent paper
\cite{Paulos:2012nu} where very similar formulas for two-loop integrals
were obtained with the help of Mellin space techniques.
These formulas generically, as mentioned previously in eq.~(\ref{mellineq}),
involve an integral over a projective space with a quadratic form in the denominator, then
integrated over one real variable (see, in particular, eq.~(8.1) of ref.~\cite{Paulos:2012nu}).  As emphasized by the authors of
ref.~\cite{Paulos:2012nu}, a great deal is known about
such projective integrals, which makes these forms particularly attractive.
As we will see below, such forms are indeed very convenient for computations.
Here, our derivation used only elementary manipulations in Feynman parameter space%
\footnote{We have found that the simple changes of variables described
above, when applied to the double-box integral
considered in eq.~(8.1) of ref.~\cite{Paulos:2012nu}, reproduces
that formula exactly, giving an elementary derivation of it.}.

\subsubsection{Step 2:  Obtaining the symbol}

Having obtained eq.~(\ref{I1v2}), we observe that several of the integrations are trivial;
for instance, we can immediately do the $b_2,b_3$ and $a_3$ integrations, leaving
\ba
 I_1 &=& \int_1^\infty dc \int \frac{d^4(a_1a_2a_\II b_1b_\II)}{\textrm{vol(GL(1))}} \frac{1}{(a_1+a_\II+b_1+b_\II)((a_1+a_\II)c+b_1+b_\II)(a_\II+b_\II+a_2\chi)}
 \nl && \hspace{5cm} \times \frac{1}{(a_1+a_2+b_1)b_\II +(b_1+a_1 c+a_2 c)a_\II} \: .  \label{I1v3}
\ea
We can easily integrate out one more variable, although this step
will unavoidably produce a logarithm. For instance, doing the
$a_2$ integration in eq. (\ref{I1v3}) yields \ba
 I_1 &=& \int_1^\infty dc \int \frac{d^3(a_1a_\II b_1b_\II)}
 {\textrm{vol(GL(1))}} \frac{\log \left(\frac{(a_\II+b_\II)(a_\II c+b_\II)}{\chi(b_\II(a_1+b_1)+a_\II(b_1+a_1 c))}\right)}{(a_1+a_\II+b_1+b_\II)((a_1+a_\II)c+b_1+b_\II)}
  \nl && \hspace{3cm}
  \times \frac{1}
  {(a_\II+b_\II)(a_\II c+b_\II)-\chi(b_\II(a_1+b_1)+a_\II(b_1+a_1 c))} \: .  \phantom{aaaa} \label{I1v4}
\ea

A pleasing surprise is that one can keep going like that, doing one integration at the time, and not hit any serious roadblock, besides the increasing length of the expressions.
For instance, $a_1$ enters in eq. (\ref{I1v4}) at most linearly in all
denominator factors and arguments of the logarithms, and so the $a_1$ integral can be done explicitly, producing dilogarithms.

If one continued like this, in the right order (doing $b_1$ and then $c$ next), the integrals produced would take the same form as eq. (\ref{I1v4}),
but with integrands involving transcendental factors of increasing transcendentality degree, for instance $\Li_3$, followed by $\Li_4$.
This is manifest from the fact that each of these integrals take the schematic form
\be
 \int_0^\infty \frac{dt}{\mbox{(linear factors in $t$)}}
 \times \big(\Li_n(\mbox{ratio of linear factors in $t$}) \hspace{0.7mm} + \hspace{0.7mm} \cdots \big) \label{typicalint}
\ee where the degree of the polylogarithm in the integrand
is $n=2$ or $3$, depending on the integral.
Integrations of this type raise the degree of transcendentality of the integrand
exactly by one. It is relatively easy to check
that the rational denominators involve only linear factors at all stages.
Namely, as a general feature of such integrals, the rational part of the measure at a given stage
is obtained simply by taking residues of the rational part of the
measure at the previous stage.
Finally, the $a_\II$ integral (setting $b_\II=1$ in this final stage)
would be of the form \be
 \frac{1}{1+\chi} \int_0^\infty \frac{da_\II}{(a_\II+1)^2} \times \big(\Li_4(\cdots) \hspace{0.7mm} + \hspace{0.7mm} \cdots \big) \label{lastint}
\ee whose measure contains a squared linear factor.  Using integration by parts
this can be replaced by a boundary term plus an integral of the form $\int
da_\II(\Li_3(\cdots)/\mbox{rational}+\cdots)$, both of which would
manifestly give rise to polylogarithms of degree 4.

Although the procedure outlined in the previous paragraphs is clearly feasible, carrying it
out in practice would be cumbersome.
It is much more efficient to perform the above integrals, exactly as described, but only at the level of the symbol.  This can be done purely algebraically and will
produce the symbol of $I_1$.  For instance, employing the algorithm described in appendix B of ref.~\cite{CaronHuot:2011kk},
each step involves only solving linear equations and can be easily automated.
The intermediate expressions are somewhat lengthy and will not be reproduced here, but in the end we obtain the very simple symbol,
\be
 \SS [I_1(\chi)] = \frac{2}{\chi} \left[\chi\otimes\chi\otimes(1+\chi)\otimes (1+\chi) \right]
  - \frac{2}{1+\chi} \left[ \chi\otimes \chi\otimes (1+\chi)\otimes \chi\right] \: . \label{symbol}
\ee
Knowing only this symbol, together with three other pieces of information, it is possible to reconstruct the function $I_1$ uniquely.

\subsubsection{Step 3: Integrating the symbol}

As we will now show, the symbol (\ref{symbol})
can be integrated unambiguously
by imposing the following three constraints on the integrated
expression:
\begin{itemize}
\item[1.] The fact that $I_1(\chi)$ has the homogeneous
transcendentality degree 4, as explained above.\footnote{A
technical issue which arises here is that, because of the
integration by parts step, lower-degree contamination is a priori possible.
This could arise if the rational
part of the measure after the integration by parts step still contains a
squared denominator $1/(a_\II+1)^2$; this would lower the transcendentality. We have verified at the level of the symbol
of the (\ref{lastint}) integrand that the integration by parts does not produce any such squared denominator.
This means that such a squared denominator could arise only from a beyond-the-symbol ambiguity in the (\ref{lastint}) integrand,
hence would have to be explicitly proportional to $\pi^2$ or $\zeta(3)$. However, our Regge asymptotics constraints
turn out to be strong enough to rule out such terms.}

\item[2.] The physical requirement that, on all physical sheets
(which can be either $-1<\chi<0$, $-\infty<\chi<-1$ or
$0<\chi<\infty$, depending on the channel under consideration),
the integral is analytic around $\chi=-1$.
This is because our integral, being planar, has a vanishing unitarity cut in the
$u$-channel and hence, by the Cutkosky rules, cannot have a
discontinuity in the $u$-channel.

\item[3.] The asymptotics (``Regge limit'') \ba
 I_1(\chi) &\to& \frac{\pi^2}{6} \log^2 \chi +
 \left(4\zeta(3)-\frac{\pi^2}{3} \right)\log \chi +\mathcal{O}(1)
 \hspace{3mm} \mbox{as} \hspace{3mm} \chi\to 0 \hspace{5mm} \mbox{and} \label{regge} \nl
 I_1(\chi) &\to& 6\zeta(3)\frac{\log \chi}{\chi} + \mathcal{O}(\chi^{-1})
 \hspace{3mm} \mbox{as} \hspace{3mm} \chi\to\infty
\ea which we have obtained directly from eq.~(\ref{I1v2}). For
instance, the double logarithm originates from the
region where $1\ll a_1\sim b_1 \ll a_2\ll 1/\chi$ (and other
variables $\sim 1$). The subleading logarithm originates from the
boundaries of that region --- explicitly, the three regions $1\sim
a_1 \ll a_2\ll 1/\chi$, $1\ll a_1 \sim  a_2\ll 1/\chi$ and $1\ll
a_1\ll a_2\sim 1/\chi$. The $\chi\to\infty$ logarithm originates from the
$a_2\to 0$ region where the other variables are $\sim 1$.
\end{itemize}

The first constraint implies that
\be
 I_1(\chi) = \frac{2}{\chi} f_1(\chi) - \frac{2}{\chi+1} f_2(\chi)  \label{I1_from_f1f2}
\ee where $f_1$ and $f_2$ are functions of homogeneous degree 4
whose symbol is consistent with eq.~(\ref{symbol})
and therefore must take the form
\ba
 f_1(\chi)&=& H_{-1,-1,0,0}(\chi) \hspace{0.8mm} + \hspace{0.6mm} \mbox{(symbol-free terms)} \: , \nl
 f_2(\chi)&=& H_{0,-1,0,0}(\chi) \hspace{0.8mm} + \hspace{0.6mm} \mbox{(symbol-free terms)} \: .
 \ea
Here ``(symbol-free terms)'' represent simpler transcendental functions multiplied by constants, such as
$\pi^2\Li_2(\cdots)$ or $\zeta(3)\log \chi$. The $H$'s are
harmonic polylogarithms \cite{Remiddi:1999ew}; explicitly, \be
 H_{-1,-1,0,0}(x) \equiv \int_0^x \frac{dt}{t+1} H_{-1,0,0}(t) \quad \mbox{and} \quad
 H_{0,-1,0,0}(x) \equiv \int_0^x \frac{dt}{t} H_{-1,0,0}(t) \: ,
\ee
with
\be
 H_{-1,0,0}(x) \equiv \frac12\int_0^x \frac{dt}{t+1} \log^2 t = -\Li_3(-x)+\log x \hspace{0.3mm} \Li_2(-x) +\frac12\log^2x\log(1+x) \: .
\ee We now impose constraint 2, regularity at $\chi=-1$.
This has to be imposed separately on $f_1$ and $f_2$ because these transcendental functions multiply different rational prefactors.
Because a function is analytic at a point if and only if its derivative is, the simplest way to proceed is to take a derivative.
For instance, near $\chi=-1$, we have
\ba
 \frac{d}{d\chi} f_1(\chi) &=& \frac{1}{\chi+1} H_{-1,0,0}(\chi) \nl
  &=& \frac{1}{\chi+1} \left( \frac{\pi^2}{6}\log \chi - \frac{\pi^2}{2}\log(1+\chi) -\zeta(3) \right)
  \hspace{0.7mm}+\hspace{0.7mm} \mbox{(terms analytic near $\chi=-1$)} \nonumber \\ \label{eq:derivative_f1}
\ea and similarly for $f_2$.
This decomposition is such that the omitted terms are
analytic near $\chi = -1$, no matter if this point is approached from either of the physical channels
$\chi \in (-\infty-i\epsilon,-1-i\epsilon)$ or $\chi \in (-1+i\epsilon,i\epsilon)$.
Thus, the non-analytic behavior of $f_1$ can be removed, simply by
subtracting the antiderivative of the $\frac{1}{\chi + 1} \big( \cdots \big)$ term in eq.~(\ref{eq:derivative_f1}),
\ba
 f_1(\chi)&=& H_{-1,-1,0,0}(\chi) -\frac{\pi^2}{6} \Li_2(-\chi) + \left(\frac{\pi^2}{4}\log (1+\chi) -\frac{\pi^2}{6}\log \chi +\zeta(3) \right) \log(1+\chi) +r_1 \: , \nl
 f_2(\chi)&=& H_{0,-1,0,0}(\chi) -\frac{\pi^2}{2}\Li_2(-\chi) + r_2 \: .
\ea
The remainders $r_1$ and $r_2$ can only be linear combinations of $\pi^2\log^2 \chi$, $\zeta(3)\log \chi$ or constants,
as they need to be devoid of branch cuts around $\chi=-1$ on all physical sheets (and at any other point other than 0 or infinity).
For instance, no dilogarithm nor any term involving $\log(\chi+1)$ would have this property. In addition, constraint 2
requires that the second term in eq.~(\ref{I1_from_f1f2}) must be pole-free at $\chi=-1$
and, therefore, we must have that $f_2(-1)=0$ on all physical branches.
Imposing this constraint will separately fix the constant term in $r_2$ and the coefficient of $\zeta(3)\log\chi$,
as the value of $\log \chi$ is equal to $\pm i \pi$, depending on the channel under consideration.

It remains to impose the Regge limits. In the case of $f_1$, the $\chi\to 0$ behavior forces $f_1(0)=0$ hence $r_1=0$.
In the case of $f_2$, the double-logarithmic term in eq.~(\ref{regge}) can be used to fix the remaining $\pi^2\log^2\chi$ ambiguity.
The function $I_1 (\chi)$ in eq.~(\ref{I1_from_f1f2}) is then uniquely fixed.  We find that the function determined this way automatically fulfills the
remaining Regge limits given in eq.~(\ref{regge}), which we view as a nontrivial consistency check.
In conclusion, we have obtained that
\ba
 f_1(\chi)&=& H_{-1,-1,0,0}(\chi) -\frac{\pi^2}{6} \Li_2(-\chi) + \left(\frac{\pi^2}{4}\log (1+\chi) -\frac{\pi^2}{6}\log \chi+\zeta(3) \right) \log(1+\chi) \: , \nl
 f_2(\chi)&=& H_{0,-1,0,0}(\chi) -\frac{\pi^2}{2}\Li_2(-\chi) - \frac{\pi^2}{12}\log^2 \chi -2\zeta(3)\log\chi - \frac{\pi^4}{20} \: .
\ea
As a cross-check on this result for $I_1 (\chi)$, we have tested it against numerical integration%
 \footnote{Using Mathematica's \texttt{NIntegrate} function to integrate the form (\ref{I1v4}).}
for a number of points with $\chi>0$ (with $\sim8$ digits precision).

Plugging this into eq.~(\ref{I++1}), we obtain the following complete results
for the chiral numerator integrals $I_{++}$ and $I_{+-}$:
\begin{eqnarray}
\hspace{-0.4cm}
 I_{++}(\chi) &=& 2H_{-1,-1,0,0}(\chi) -\frac{\pi^2}{3}\Li_2(-\chi) + \left(\frac{\pi^2}{2}\log (1+\chi) -\frac{\pi^2}{3}\log \chi+2\zeta(3) \right) \log(1+\chi) \nl &&
 - \hspace{0.6mm} 6 \chi \zeta(3) \: ,\label{eq:chiral_DBs_final_result_1}
 \\
\hspace{-0.4cm}
 I_{+-}(\chi) &=& 2H_{0,-1,0,0}(\chi) -\pi^2\Li_2(-\chi) - \frac{\pi^2}{6}\log^2 \chi -4\zeta(3)\log\chi - \frac{\pi^4}{10}-6(1+\chi)\zeta(3) \: . \phantom{aaa} \label{eq:chiral_DBs_final_result_2}
\end{eqnarray}
These formulas are such that, with the standard branch choice for the polylogarithms, the result is real in the Euclidean region $\chi>0$.
Also, we refer to the footnote around eq.~(\ref{Feynman}) for an explanation of our conventions.
If desired, these results could be rewritten in terms of classical polylogarithms such as $\Li_4$, but
we have not found such rewritings particularly illuminating.

Equations (\ref{eq:chiral_DBs_final_result_1})-(\ref{eq:chiral_DBs_final_result_2}) contain $\zeta(3)$
terms which violate the uniform transcendentality degree of the other terms, which may be surprising at first sight.
We tentatively attribute this to double-triangle integrals present in the difference between our $I_{+ \pm}$ and the ``true'' chiral integrals,
as discussed in the footnote above eqs.~(\ref{def_chiral_integrals_1})-(\ref{def_chiral_integrals_2}).
It would be interesting to evaluate explicitly the difference and see if the $\zeta(3)$-terms disappear.

As we were hoping, we have found that the chiral numerator integrals $I_{+ \pm}$ admit rather compact analytical expressions.
We invite the reader to compare our results against earlier results
in the literature for double-box integrals. We refer to eqs. (22)-(25) of ref. \cite{Smirnov:1999gc}
for the analytical result for the scalar double box, and to eq. (13) of ref. \cite{Anastasiou:2000kp} for
the analytical result for the double box with the $(\ell_1 + k_4)^2$ numerator insertion.

\section{Discussion and conclusions}\label{sec:conclusions}

Generalized unitarity is a method for computing loop-level scattering amplitudes
that has been applied very successfully at one loop,
in particular to computations of processes with many partons in the final state.
In ref.~\cite{Kosower:2011ty}, the first steps were taken in
developing a fully systematic version of generalized unitarity at two loops.
In the approach followed there,
the two-loop amplitude is decomposed as a linear combination of basis
integrals, in similarity with eq.~(\ref{BasicEquation}). The goal
of the calculation is the determination of the integral coefficients
as functions of the external momenta; once this is done, the amplitude
is determined.  At two loops, the integrals with the leading topology
in the basis decomposition have the double-box topology, illustrated in
figure~\ref{fig:general_double_box}.\footnote{When the number of external
states exceeds four, the leading topology is that of a pentagon-box or a double-pentagon.
However, we expect the coefficients of such integrals to be simpler to extract due to the explicit octacuts they contain.}
The integral coefficients of the double-box integrals are determined by applying
to both sides of the basis decomposition of the two-loop amplitude so-called
augmented heptacuts. These are defined by replacing the seven propagators in
the double-box integrand by complex $\delta$-functions. This will freeze seven
out of the eight degrees of freedom in the two loop momenta. The Jacobian
arising from solving the $\delta$-function constraints contains poles, known as
leading singularities, and the remaining integration can be chosen as a contour in the complex plane, enclosing these
leading singularities.

Our strategy towards applying the generalized-unitarity method to two-loop QCD amplitudes is not to try to determine
the complete QCD integrand, but only the coefficients of a small (possibly minimal) number of craftily chosen ``master'' integrals,
to which all the other ones can be reduced using integral identities (for example, integration-by-parts identities or Gram determinant identities).
Projecting out these identities imposes constraints on the contours, namely, that on allowed contours the integral identities which are valid on
the physical (uncut) contour must remain valid.  As explained in ref.~\cite{Kosower:2011ty},
contours satisfying this consistency condition are guaranteed to produce correct results for scattering amplitudes in any gauge theory.\footnote{We stress that the method makes no assumption regarding the powers of loop momentum present in numerators. Thus, as discussed at the beginning of
section~\ref{sec:proliferation_of_contours}, our method applies indiscriminately to the contributions from planar double boxes in any quantum field theory.}
These \emph{master contours} can be chosen so as to extract the coefficient of any particular master integral in a given basis.
A perplexing feature of the master contours obtained in ref.~\cite{Kosower:2011ty}
is that they are not uniquely defined: indeed, they were found to be characterized by 6 free parameters.
\\
\\
\noindent
In this paper, we explain this phenomenon as a simple redundancy of variables and
find that the two-loop master contours are unique, in perfect analogy
with the situation in one-loop generalized unitarity.

Our starting point is a careful examination of the solutions to the heptacut
of the double-box integral with an arbitrary number of external states.
As mentioned above, the heptacut amounts to
setting all the propagators in the double-box graph on-shell. This
will freeze all but one of the degrees of freedom in the two loop momenta;
for generic external momenta, this degree of freedom is necessarily complex
and thus naturally parametrizes a Riemann surface. We have provided
a complete classification of the solutions to the heptacut constraints,
explained in section~\ref{sec:solutions}, based on the number of three-point vertices in the double-box graph.
We find that as the number of three-point vertices is decreased,
there are six, four or two Riemann spheres, intersecting
pointwise and linked into a chain; or, ultimately, an elliptic curve.
We find that the intersection points of these spheres coincide
with the poles of the Jacobian arising from linearizing
the heptacut constraints; i.e., the two-loop composite leading singularities.
In section \ref{sec:proliferation_of_contours} we explain that this gives rise to identifications which explain the mystery found in ref.~\cite{Kosower:2011ty}.

Moreover, we find that at the intersection points of the Riemann spheres, one of the two
loop momenta becomes collinear with the massless external
momenta attached to the respective vertices of the double-box
graph. As integration regions where the loop momentum is becoming
collinear with massless external momenta are associated
with infrared divergences in the uncut loop integral, this
provides a natural physical interpretation of two-loop
leading singularities.

In the case when the double-box graph contains no three-point vertices,
we find that the Riemann surface associated with the maximal cut is an elliptic curve; i.e., a torus.
As discussed in subsection \ref{sec:elliptic}, the presence of an elliptic curve
opens up a possibility to connect, for the first time precisely,
the long-held belief that the maximal cuts of a given integral should be connected to
the analytic structure of its integrated expression.
In particular, we argued that an integral whose heptacut contains an elliptic curve is unlikely to be
expressible in terms of polylogarithms.

The integral coefficients of two-loop amplitudes are functions of the dimensional regulator $\epsilon$.
The extraction of the $\mathcal{O}(\epsilon)$ contributions poses a technical difficulty
as they may multiply $\frac{1}{\epsilon^k}$ singularities in the integrated
expressions for the two-loop integrals and thus produce finite contributions
to the amplitude. Unfortunately, they are not obtainable from
cuts performed in strictly four dimensions and must instead be computed
by taking cuts $D=4-2\epsilon$ dimensions, something which is technically much more involved.
One potential way to minimize this technical
problem would be to use a basis in which as many integrals as possible are
infrared finite. Namely, although the integral coefficients may still have
$\mathcal{O}(\epsilon)$ corrections, these contributions would then multiply
infrared finite integrals and hence would not be physically relevant.
Of course, as two-loop amplitudes do have infrared divergences,
a basis of integrals must necessarily contain some infrared divergent integrals,
but it is still plausible that a basis containing as few such integrals
as possible will minimize the amount of work needed to obtain their coefficients.
In section~\ref{sec:proliferation_of_contours} we have shown that the chiral integrals
recently introduced by Arkani-Hamed et al. provide a basis for the double-box
contributions to (four-point) two-loop amplitudes in any gauge theory. These integrals
are infrared finite, allowing for their coefficients to be obtained from
strictly four-dimensional cuts.

In section~\ref{sec:analytical_evaluation_of_chiral_DBs} we evaluated
these integrals analytically for the case of four massless external momenta.
As hoped, we have found that they are given by very simple and compact expressions
 (see eqs.~(\ref{eq:chiral_DBs_final_result_1})-(\ref{eq:chiral_DBs_final_result_2}) for our final results).
Our computation consisted of a more or less direct Feynman parameter integration, together with a few tricks.
It will be very interesting to see if a similar finite basis of masters is available in the case of five particles, in the double-box and pentagon-box topologies.
Moreover, we are very hopeful that such finite integrals can be computed analytically by some means.

Using the integral identities which have been explicitly constructed at five points \cite{Gluza:2010ws},
we believe it will be possible to explicitly construct the (unique) master contours in this case using our results. We hope to return to this question in the near future.
In general, a better understanding of integration-by-parts and Gram determinant identities would be very helpful.
Although in this work we have not considered other topologies with seven or more propagators, such as pentagon-boxes,
we expect the extraction of their coefficients to be much simpler than that of the double boxes, due to the presence of explicit octacuts.
It will also be important to understand their extraction in the future.
In addition, it is important to understand the extraction of coefficients of integrals with a smaller number of propagators.
Other natural extensions of our work would include non-planar topologies, or integrals with internal masses.
We are hopeful that the ideas presented in this paper will be useful for answering these questions.

Another important question concerns the dimension of the space of master integrals of a given topology.
Although for this particular question we have only considered the case of four particles,
we conjecture that in the general case, the master contours are still unique, and that
their number is precisely equal to the number of independent master integrals of the corresponding double-box topology.
Verifying this would be very interesting.

\acknowledgments

We have benefited from discussions with Nima Arkani-Hamed, Henrik
Johansson, David Kosower, Donal O'Connell and Edward Witten.
SCH gratefully acknowledges support from the Marvin L.~Goldberger Membership and from
the National Science Foundation under grant PHY-0969448. 
KJL gratefully acknowledges financial support from the
European Program Advanced Particle Phenomenology in the LHC
Era under contract PITN-GA-2010-264564 (LHCPhenoNet).
This work is supported by the European Research Council under Advanced
Investigator Grant ERC--AdG--228301.

\begin{appendix}

\section{Residues of maximally cut amplitudes}\label{sec:cuts_across_kin_sols}

An amusing application of the identities
(\ref{eq:residues_of_cuts_relation_1})-(\ref{eq:residues_of_cuts_relation_6})
arises in the context of taking residues at Jacobian poles of the
heptacut two-loop amplitude $\left. J(z) \prod_{j=1}^6
A_j^\mathrm{tree} (z) \right|_{\mathcal{S}_i}$. In
generalized-unitarity calculations, this quantity forms the input
out of which the integral coefficients of the two-loop amplitude
are computed. The identities
(\ref{eq:residues_of_cuts_relation_1})-(\ref{eq:residues_of_cuts_relation_6})
relate, in particular, the residues of the heptacut two-loop
amplitude across different kinematical solutions. As a result,
they explain the vanishing of certain residues, as well as the
seemingly accidental equality between pairs of other residues.
This in turn allows one to cut the work of evaluating these
residues in half.

As an example, let us consider the heptacut
illustrated in figure~\ref{fig:5-point_turtle_box_S_3,4,5} of the
two-loop amplitude $A^{(2)}(1^-, 2^-, 3^+, 4^+, 5^+)$. The helicities
assigned to the internal and external states allow only gluons to
propagate in the loops; this in turn implies \cite{Bern:2009xq} that the results
for the heptacut amplitude within $\mathcal{N}=4,2,1,0$ Yang-Mills theory
are identical.

\begin{figure}[!h]
\begin{center}
\includegraphics[angle=0, width=0.45\textwidth]{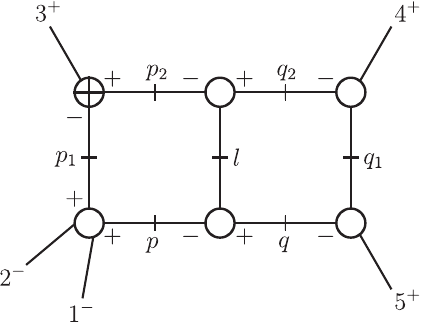}
\caption{A heptacut of the two-loop amplitude $A^{(2)}(1^-, 2^-,
3^+, 4^+, 5^+)$. The heptacut two-loop amplitude $\left. J(z) \prod_{j=1}^6
A_j^\mathrm{tree} (z) \right|_{\mathcal{S}_i}$ only receives
nonvanishing contributions from kinematical solutions consistent
with the assigned internal helicities.}
\label{fig:5-point_turtle_box_S_3,4,5}
\end{center}
\end{figure}

The heptacut $\left. J(z) \prod_{j=1}^6
A_j^\mathrm{tree} (z) \right|_{\mathcal{S}_i}$ shown in
figure~\ref{fig:5-point_turtle_box_S_3,4,5} of this amplitude only receives
nonvanishing contributions from kinematical solutions consistent
with the internal helicites shown there --  in particular
kinematical solutions whose $(3, p_1, p_2)$-vertex is
$\overline{\mathrm{MHV}}$. By inspection of figure~\ref{fig:Jacobian_poles_1} we thus see that
\begin{equation}
\left. \prod_{j=1}^6 A_j^\mathrm{tree} (z) \right|_{\mathcal{S}_i}
\hspace{1mm}=\hspace{1mm} 0 \hspace{6mm} \mbox{for} \hspace{3mm}
i=1,2,6 \: . \label{eq:cut_amp_vanishes_on_S_1,2,6}
\end{equation}
For the heptacut amplitude evaluated at the remaining three kinematical
solutions, direct calculation reveals the residues at the
Jacobian poles to be
\begin{eqnarray}
\frac{1}{i A^\mathrm{tree}_{--+++}} \left.
\mathop{\mathrm{Res}}_{z=(0, \hspace{0.4mm} Q_1)} J(z) \prod_{j=1}^6 A_j^\mathrm{tree} (z)
\right|_{\mathcal{S}_3} \hspace{2mm}&=&\hspace{1mm} \left(1, \hspace{0.7mm}
-1 \right) \\
\frac{1}{i A^\mathrm{tree}_{--+++}} \left.
\mathop{\mathrm{Res}}_{z=(0, \hspace{0.4mm} P_1)} J(z) \prod_{j=1}^6 A_j^\mathrm{tree} (z)
\right|_{\mathcal{S}_4} \hspace{2mm}&=&\hspace{1mm} \left(0, \hspace{0.7mm}
1 \right) \\
\frac{1}{iA^\mathrm{tree}_{--+++}} \left.
\mathop{\mathrm{Res}}_{z=(0, \hspace{0.4mm} P_1)} J(z) \prod_{j=1}^6 A_j^\mathrm{tree} (z)
\right|_{\mathcal{S}_5} \hspace{2mm}&=&\hspace{1mm} \left(0, \hspace{0.7mm}
-1 \right) \: .
\end{eqnarray}
One observes that the residues at $z=0$ in solutions $\mathcal{S}_4$ and $\mathcal{S}_5$
are vanishing. This can now be easily explained by eqs.
(\ref{eq:residues_of_cuts_relation_5}) and (\ref{eq:residues_of_cuts_relation_2}) as a consequence
of the vanishing (\ref{eq:cut_amp_vanishes_on_S_1,2,6})
of the heptacut amplitude on solutions $\mathcal{S}_6$ and $\mathcal{S}_2$, respectively.
Moreover, eq.~(\ref{eq:residues_of_cuts_relation_4}) relates the residues at the nonzero Jacobian poles in $\mathcal{S}_3$
and $\mathcal{S}_4$; similarly, eq. (\ref{eq:residues_of_cuts_relation_3})
relates residues in $\mathcal{S}_3$ and $\mathcal{S}_5$. In conclusion, the identities
(\ref{eq:residues_of_cuts_relation_1})-(\ref{eq:residues_of_cuts_relation_6})
allow us to cut the work of evaluating the residues of the heptacut two-loop
amplitude at the Jacobian poles in half.

\section{An elliptic curve in planar $\mathcal{N}=4$ super Yang-Mills}\label{app:N4}

Computations in $\mathcal{N}=4$ super Yang-Mills (SYM) theory are generally much easier than in QCD or pure Yang-Mills,
in no small part due to its so-called dual conformal (super)symmetry, which is a hidden symmetry of the planar limit of the
theory, invisible from its Lagrangian \cite{Drummond:2008vq}.
Several two-loop amplitudes have been computed analytically so far, and all have been expressible in terms of special functions called degree-4 (multiple) polylogarithms.
In some cases, using this information, it was even possible to guess nontrivial amplitudes \cite{Dixon:2011pw,Dixon:2011nj}.
Therefore, it is an important question whether all amplitudes in planar $\mathcal{N}=4$ SYM are given by polylogarithms.

Here we wish to give evidence that $\mathcal{N}=4$ SYM knows about
much more than polylogarithms, even when only massless internal states are present.  We
will do so by exhibiting a specific helicity configuration for
10 particles which will be given by the single integral in
figure~\ref{fig:elliptic10}, plus nothing else, ruling out any possible cancelations. As argued in the main text, we
find it extremely unlikely that this integral is expressible in terms of polylogarithms.

We claim that the 10-scalar N${}^3$MHV amplitude in $\mathcal{N}=4$ SYM with the SU(4)${}_R$-symmetry assignment $h=(\phi^{12},\phi^{23},\phi^{23},\phi^{34},\phi^{34},\phi^{34},\phi^{41},\phi^{41},\phi^{12},\phi^{12})$ is equal to
\begin{equation}
A_{10,h} = g^4 I_{2,2,1,2,2,1}(p_1{+}p_2, \hspace{0.6mm} p_3{+}p_4, \hspace{0.6mm}
p_5, \hspace{0.6mm} p_6{+}p_7, \hspace{0.6mm} p_8{+}p_9, \hspace{0.6mm} p_{10}) + \mathcal{O}(g^6)  \label{Amp10}
\end{equation}
where $g^2\equiv\frac{g^2_\textrm{YM}N_c}{16\pi^2}$.

\begin{figure}[!h]
\begin{center}
\includegraphics[angle=0, width=0.45\textwidth]{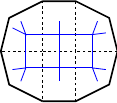}
\caption{(Color online). A decagonal Wilson loop, with the dashed
lines representing the seven scalar propagators in the unique Feynman
diagram contributing to the Grassmann component described in the main text.  We
have superimposed in blue its dual graph, the double box, which is
given by the same integral.}\label{fig:wilsonloop}
\end{center}
\end{figure}

The argument uses the amplitude/(super-)Wilson loop duality \cite{
Alday:2007hr,Drummond:2007aua,Drummond:2008aq,Bern:2008ap,Mason:2010yk,CaronHuot:2010ek}.
We try to review here the essential information.  The helicity information, from the Wilson loop viewpoint, is most usefully encoded into Grassmann variables
$\chi_i^A$ ($A=1,\ldots, 4$) transforming in the ${\bf 4}$ of SU(4)${}_R$.
Using the dictionary in ref.~\cite{Hodges:2009hk}, the helicity configuration $h$ thus maps to the (momentum super-twistor)
component $(\chi_1^2\chi_2^2\chi_3^3\chi_4^3\chi_5^3\chi_5^4\chi_6^4\chi_7^4\chi_8^1\chi_9^1\chi_{10}^1\chi_{10}^2)$ of the super Wilson-loop.
Notice that 12 $\chi$'s are turned on, which is the correct number for an N${}^3$MHV amplitude.

In ref.~\cite{CaronHuot:2010ek}, the super Wilson loop was expressed in the form
\begin{equation}
 \langle W_{10,\chi} \rangle = \prod_{i=1}^{10} \mathcal{V}_i \mathcal{E}_i, \quad\mbox{where}\quad
%\end{equation}
%where
%\begin{equation}
 \mathcal{V}_i = \frac{\phi_{AB}\chi_i^A\chi_{i{+}1}^B}{\langle k_i|k_{i{+}1}\rangle} + C \chi_i\chi_i + \mathcal{O}(\chi^3) \label{vertex}
\end{equation}
are vertex factors, and the edges are $\mathcal{E}_i=\mathcal{P} e^{-i \int A_\mu dx^\mu + \mathcal{O}(\chi)}$.
The $k_i$ are the on-shell momenta of the scattering amplitudes, which are also the lengths of the Wilson polygon segments.
As we will now argue, the chosen Grassmannian component is cooked up to make the edge factors 1 at this order; that is, $\mathcal{E}_i\to 1$.
To see this, we will show that for this component there is no coupling to the fermions of the theory.
This will rely on very general properties of the edge interactions, which is why we do not give explicit expressions here.
For instance, the edge $\mathcal{E}_i$ contains a coupling of the form $\psi \chi_i\chi_i\chi_i$, which vanishes here because no cubic power of any given $\chi_i$ is present in the Grassmann component under consideration. The vertex factor between $k_i$ and $k_{i{+}1}$ contains cubic interactions of the form $\chi_i\chi_i\chi_{i{+}1}$, but no such product can form a ${\bf 4}$ of SU(4)${}_R$ for the chosen component.  We conclude that all the couplings of the Wilson loop to $\psi$ vanish.  Direct couplings to the field strength tensor $F_{\mu\nu}$ vanish for the same reasons.
Regarding the couplings to $\bar\psi \chi$, these come from edge integrals of the form $\mathcal{E}\propto \int \bar\psi \chi$.  If such a coupling were to contribute,
the $\bar\psi$ propagator would have to land on a $\psi$ field, which could only come from a Lagrangian interaction term of the form $g_\textrm{YM}\psi\psi\phi$.
While this interaction exists, it is too inefficient a way to use one power of $g_\textrm{YM}$, for an amplitude which must couple 12 essentially distinct $\chi$'s using only four powers of $g_\textrm{YM}$ (it could only contribute at order $g_\textrm{YM}^6$).

We conclude that with the chosen component, the Wilson loop couples only to the scalars of the theory.
On the edge corresponding to $k_i$ there is a $\phi\chi_i\chi_i$ coupling, which could in principle contribute for edges $5$ and $10$ because our component does contain two $\chi_5$ and two $\chi_{10}$.   But this would require the next two $\chi$'s, for instance $\chi_{6}^4$ and $\chi_7^4$, to also couple to a scalar.
However, this would not be allowed by the SU(4)${}_R$ symmetry.  We conclude that the edge factors $\mathcal{E}_i$ are trivial for the chosen
Grassmannian component,  and that only the vertex factors shown in eq.~(\ref{vertex}) contribute.

Thus, the expectation value of the supersymmetric Wilson loop, for the chosen component, reduces to a correlation function of 6 scalars, each of which couples to two $\chi$ Grassmann components.  The six scalars in turn propagate to two $g^2_\textrm{YM}[\phi,\phi]^2$ interactions from the Lagrangian, producing a contribution of order $g^4_\textrm{YM}$.  All possible distributions of the scalars among the vertices have to be considered, but for the chosen component it is possible to see that only one distribution is allowed by the SU(4)${}_R$ symmetry.  It gives rise to the Feynman diagram shown in figure~\ref{fig:wilsonloop}; this is the one and only diagram which contributes.
The seven scalar propagators in the diagram reproduce precisely the seven denominators in the right hand side of
eq.~(\ref{Amp10}) --- the Feynman graphs are dual to each other. That is, the momentum space expression for one is equal to the configuration space expression for the other.
We omit the verification that the prefactor works out as claimed.

This agreement concerns the integrand in exactly four dimensions.  However, the chosen component is infrared
finite at two loops because, by infrared exponentiation theorems, any IR divergence of
the two-loop amplitude would have to be canceled by IR divergences in the tree and one-loop amplitudes.
But as the latter vanish (in $D$ dimensions) for this process, the two-loop amplitude is necessarily IR
finite. For this reason, it is certainly enough to consider only the four-dimensional integrand in this case.
We have also checked the present result for the four-dimensional integrand against a computer implementation of the recursion relation for the integrand
\cite{ArkaniHamed:2010kv}, finding perfect agreement.

\end{appendix}

\providecommand{\href}[2]{#2}\begingroup\raggedright\endgroup

\end{document}